\documentclass[12pt]{article}
\usepackage[explicit]{titlesec}
\usepackage[none]{hyphenat}
\usepackage{amsmath,amssymb,amsfonts,mathrsfs}

\usepackage{soul}
\setul{.6pt}{.4pt}

\usepackage{geometry}
\usepackage{fancyhdr}
\usepackage[labelfont={footnotesize,bf} , textfont=footnotesize]{caption}
\titleformat{\section}
{\normalfont\bfseries}{\thesection.}{0.5em}{\MakeUppercase{\textbf{#1}}}
\titlespacing\section{0pt}{0pt}{-10pt}
\titleformat{\subsection}
{\normalfont\itshape}{\thesubsection.}{0.5em}{\textit{#1}}
\titlespacing\subsection{0pt}{0pt}{-8pt}

\makeatletter
\newcommand\sixteen{\@setfontsize\sixteen{17pt}{6}}
\renewcommand{\maketitle}{\bgroup\setlength{\parindent}{0pt}
	\begin{flushleft}
		\sixteen\bfseries \@title
		\medskip
	\end{flushleft}
	\textit{\@author}
	\egroup}
\makeatother

\usepackage{caption}
\usepackage{graphicx,subfigure}
\graphicspath{{}}

\usepackage{booktabs}
\usepackage{multirow}
\usepackage{array}

\usepackage[american]{babel}
\usepackage{microtype} 
\usepackage{enumerate}

\title{Molecular dynamics simulation of crack growth in mono-crystal nickel with voids and inclusions}

\author{
	Zhenxing Cheng$^{a}$, Hu Wang$^{a*}$, Gui-Rong Liu$^{b}$, Guangyao Li$^{a}$ \medskip \\
	$^{a}$State Key Laboratory of Advanced Design and Manufacturing for Vehicle Body, Hunan University, Changsha, 410082, PR China \\
	$^{b}$Department of Aerospace Engineering and Engineering Mechanics, University of Cincinnati, Cincinnati, Ohio, 45221, United States\\}

\begin{document}
	
	\vspace*{.01 in}
	\maketitle
	\vspace{.12 in}
	
	\section*{abstract}
	In this study, the crack propagation of the pre-cracked mono-crystal nickel with the voids and inclusions has been investigated by molecular dynamics simulations. Different sizes of voids, inclusions and materials of inclusions are used to fully study the effect of the voids and inclusions during the crack propagation process. The dislocations evolution, stress distribution and crack length are analyzed as the associated mechanical properties. The results indicate that the voids and inclusions can change the path of crack propagation of the pre-cracked mono-crystal nickel. Moreover, the results show that the voids and inclusions can lead a better resistance to plastic deformation of the mono-crystal and the inclusions can make the system more difficult to fracture.
	
	\textit{Keywords}: Molecular dynamics, Crack propagation, Crystals, Dislocations
	
	\vspace{.12 in}
	
	
	\section{introduction}
	
	Damage and fracture of engineering materials is a critical issue in engineering fields due to the effect on the stability of structures and components. Generally, the fracture process includes two parts, named crack initiation and propagation. The propagation of a crack is associated with the defects in the specimen, including dislocations, vacancies, slip bands, twins, voids and inclusions. In view of these facts, it is necessary to understand the crack propagation on the basis of different length scales, e.g., macro-scale, micro-scale, nano-scale (or atomic scale) \cite{benz2015reconsiderations,li2015fatigue,huang2012discrete,leung2014atomistic,zhang2017mechanisms,yasbolaghi2020micro}. On the macroscopic scale, many numerical methods have been developed to study the crack propagation, including the finite element method (FEM) \cite{branco2015review}, boundary element method (BEM) \cite{santana2016dual}, meshless method \cite{khosravifard2017accurate}, extended finite element method (X-FEM) \cite{belytschko2009review,cheng2019exact}. On the microscopic scale, crystal plasticity finite element method (CPFEM) is a popular method to analyze the plastic deformation at the crack tip in poly-crystals \cite{li2014crystal,zhang2018quantitative,proudhon20163d}. On the atomic scale, molecular dynamics (MD) method has been widely used to study material fracture behaviors.
	
	During the last couple of years, MD simulation is emerging as a widely used method to investigate the behavior of crack tips in different types of crystalline materials. Tang et al. employed MD method to study the behavior of the crack tip in magnesium single crystal under cyclic loading \cite{tang2010fatigue}. Horstemeyer et al. reviewed the research related to atomic simulations of fatigue crack propagation in FCC metals \cite{horstemeyer2010nanostructurally}. Borodin and Vladimirov used MD simulations to understand the three-dimensional kinetic of micro-crack propagation in $\alpha$-iron and the accompanying lattice transformations at the crack tips \cite{borodin2011molecular}. Wu and Yao also studied the edge crack growth, the micro-structure evolution and the stress distribution in a single crystal of nickel by MD based simulations \cite{wu2012molecular}. Then Ma et al. studied the micro-structure evolution of fatigue crack growth in BCC iron at 300K by MD \cite{ma2014molecular} and Cui et al. studied the edge crack behavior in single crystals of aluminum and copper under mode \uppercase\expandafter{\romannumeral1} loading conditions \cite{cui2014molecular}. Subsequently, studies of the crack and void growth in a single crystal of nickel have been published, including the edge crack growth, the central crack growth, the fatigue crack growth \cite{wu2015molecular,li2015cohesive,sung2015studies}. After that, Zhang et al. studied the mechanisms of crack propagation in single crystal, bicrystal and tricrystal of nickel by MD simulations \cite{zhang2017mechanisms}. As for some other materials, Chandra et al. used MD to study the crack growth behavior in aluminum in the presence of vacancies \cite{chandra2016molecular} and Fang et al. studied the grain boundary geometry on the crack propagation of bi-crystal aluminum \cite{fang2016molecular}. Feng et al. used MD simulations to investigate the orientation and temperature dependence of tensile response in single crystal W \cite{feng2018twin}. Chowdhury et al. studied the fracture mechanism and the effects of surface crack on the mechanical properties of silica glass by MD \cite{chowdhury2019effects} and Lu et al. simulated the crack propagation in poly-crystalline NiTi alloys \cite{lu2020cohesive}.
	
	To our knowledge, few writing works have reported the mechanism of crack propagation in crystalline metal materials with void or inclusion. In particular, the effect of voids and inclusions on crack propagation has never been reported on the microscopic scale while some literature indicated that voids and inclusions could change the path of crack propagation at the macroscopic scale \cite{giner2009abaqus,cheng2018control,jiang2020cpot}. In this study, the mechanical properties of the crack propagation for pre-cracked mono-crystals with voids and inclusions have been investigated by MD simulations. Moreover, the effect of voids and inclusions on the mechanical properties of crack propagation has been studied and the inclusions with different materials have also been considered.
	
	The remainder of this study is organized as follows. The basic theory of MD and the atomic model for the crack are introduced in section 2. The results and discussions of MD simulations can be found in section 3. Finally, some conclusions are summarized in section 4.
	
	\section{Methodology}
	
	\subsection{Basic theories of MD}
	
	In this work, MD simulations of a pre-cracked mono-crystal with different voids and inclusions were performed to investigate the effect of voids and inclusions on crack propagation. Generally, MD simulations can be described by the position and momentum of each atom or molecule in a simulation box. The dynamics of atoms follow the Newton's second law which can be described as the following equation:
	
	\begin{equation} \label{eq:1}
	m_i\vec{a_i} = \vec{f_i} = -\triangledown U({\vec{r_i}}),
	\end{equation}
	
	\noindent where $m_i$ and $\vec{a_i}$ are the mass and acceleration of atom $i$ respectively, $\vec{r_i}$ means the position of atom $i$ and $U$ denotes the inter-atomic potential. A typical example of the inter-atomic potential is the Embedded Atom Method (EAM) in which an atom should be regarded as the embedded component of a lattice \cite{daw1983semiempirical}. Specially, the EAM potential can be written as
	
	\begin{gather}
	U_{ij} = \sum_i F_i\left( \rho_i \right)  + \frac{1}{2}\sum_i \sum_{j\neq i} \phi_{ij}\left( r_{ij}\right), \\
	\rho_i = \sum_{j\neq i} \rho_j \left( r_{ij}\right) , \label{eq:3}
	\end{gather}
	
	\noindent where $U_{ij}$ is the potential function to describe the interaction between atom $i$ and $j$. The symbol $F_i$ means the embedding energy depended on the electron cloud density $\rho_i$ around the atom $i$. The electron density $\rho_i$ is associated with all the atoms in the system which can be calculated by Equation\eqref{eq:3}. The symbol $\phi_{ij}$ denotes the pairwise potential, which depends on the relative distance $r_{ij}$ between atom $i$ and its neighbor atom $j$. Generally, most empirical potentials can be written as
	
	\begin{equation}
	U(\vec{r_i})=\sum_i U_{ij} (\vec{r_i},\vec{r_j}),
	\end{equation}
	
	\noindent where $U$ is a function of the energy of each atom ($U_{ij}$), which depends on the position of atoms $i$ and $j$. Usually, atoms do not interact directly beyond the cut-off radius $r_{cut}$, which implies that
	
	\begin{equation}
	\vec{f_i} = -\triangledown U({\vec{r_i}}) = 0, \quad if \quad r_{ij}>r_{cut}.
	\end{equation}
	
	There are a series of motion equations needs to be solved during the MD simulation, such as Equation\eqref{eq:1}. Thus, the Velocity-Verlet algorithm is employed as the time integration algorithm to solve the motion equations with a considerable accuracy. More details about the Velocity-Verlet algorithm can be found from the reference\cite{omelyan2002optimized}.

	The definition of stress for an atomic simulation is different from the continuum stress concept. A common definition of virial stress suggested by Swenson et al. \cite{swenson1983comments} is used in this study. Atomic scale virial stresses are equivalent to the continuum Cauchy stresses \cite{subramaniyan2008continuum}. The stress contains two parts, potential and kinetic energy parts, which is defined as
	
	\begin{equation}
	\sigma_{xy}=\frac{1}{v^i}\sum_i \left[ \frac{1}{2}\sum_{j = 1}^N \left( r_x^j - r_x^i \right) f_y^{ij}-m^i v_x^i v_y^i \right] ,
	\end{equation}
	
	\noindent where $m^i$ means the mass of atom $i$, the subscripts $x$ and $y$ denote the Cartesian components and $v^i$ is the volume of the atom $i$. The superscripts $i$ and $j$ are the atom identification number, which mean atom $i$ and $j$. The symbols $r$, $f$ and $v$ denote the relative position, inter-atomic force and velocity respectively. Specially, the symbol $f_y^{ij}$ is the $y$ direction force on atom $i$ induced by atom $j$, $v_x^i$, $r_x^i$ are the velocities and relative position of atom $i$ along the $x$ direction. Other symbols are as similar as the above. To roughly calculate the local stress field of the system, the atomic stress $\sigma_{xy}$ for each atom in the system is used to plot the stress contours. Here, the atomic stress $\sigma_{xy}$ has the unit of stress$\times$volume. Then the Von Mises stress $\bar{\sigma}$ can be calculated by
	
	\begin{scriptsize}
		\begin{equation}\label{eq:8}
		\bar{\sigma}=\frac{1}{\sqrt{2}}\sqrt{(\sigma_x-\sigma_y)^2+(\sigma_y-\sigma_z)^2+(\sigma_z-\sigma_x)^2+6(\tau_{xy}^2+\tau_{yz}^2+\tau_{zx}^2)},
		\end{equation}
	\end{scriptsize}
	
	\noindent where $\sigma_x, \sigma_y, \sigma_z$ are the normal stresses and $\tau_{xy}, \tau_{yz}, \tau_{zx}$ are the tangential stresses.

	\subsection{Atomic model}
	
\begin{figure*}[htb]
		\centering
		\subfigure[Mono-crystal]{\includegraphics[width=1.45in]{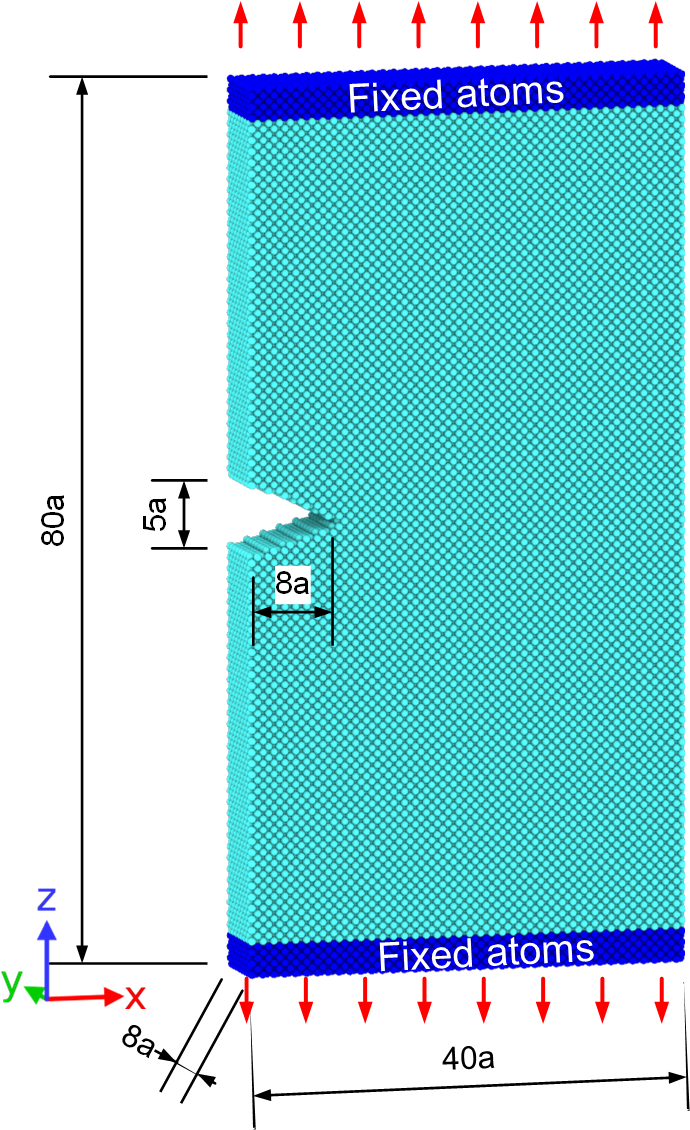}}
		\hspace{0.1in}
		\subfigure[With a void]{\includegraphics[width=1.4in]{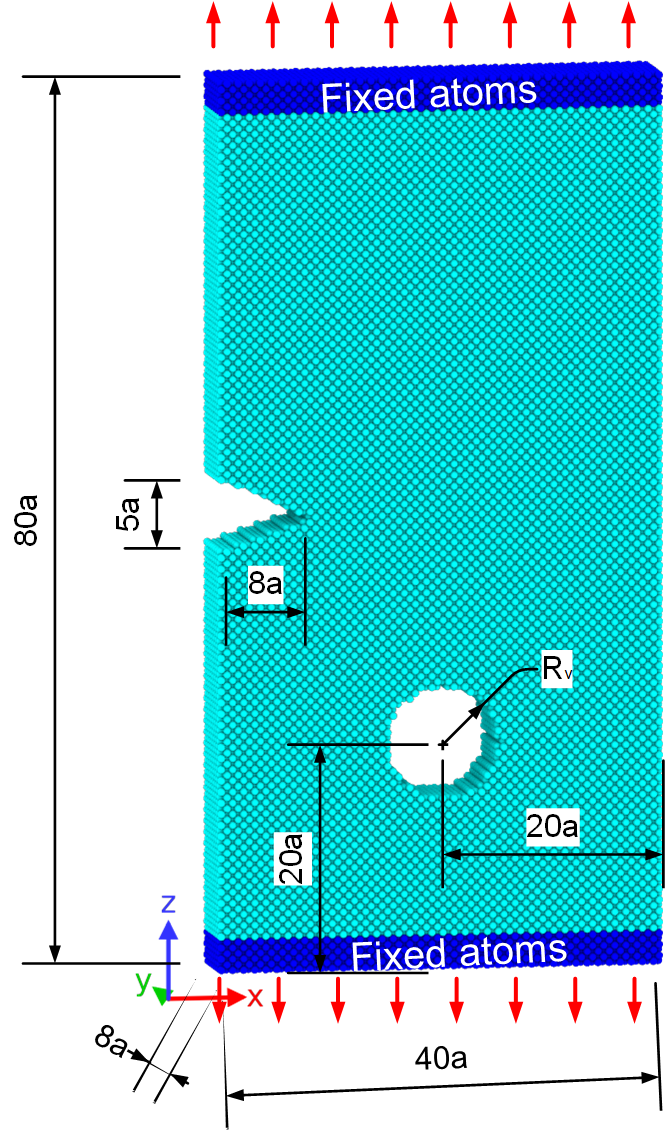}}
		\hspace{0.1in}
		\subfigure[With a inclusion]{\includegraphics[width=1.4in]{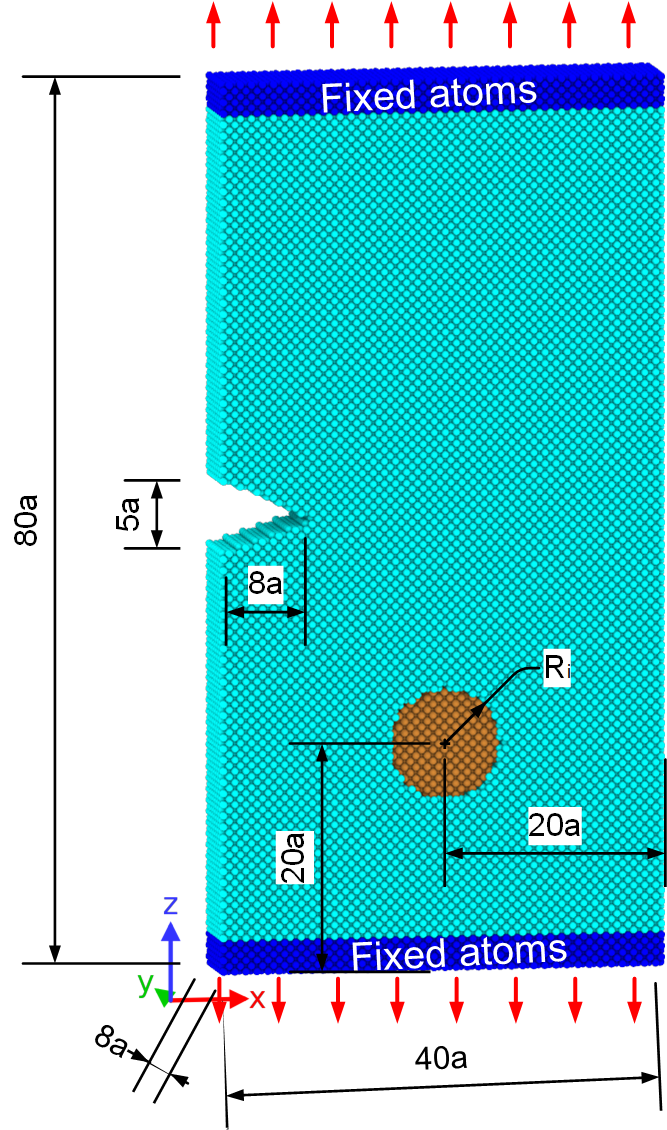}}
		\caption{Atomic models for the crystal crack propagation.}
		\label{fig:model}
	\end{figure*}

	Figure \ref{fig:model}a shows a three-dimensional MD atomic model for the crack propagation under the mode \uppercase\expandafter{\romannumeral1} loading conditions. The material of the model is nickel with face center cubic (FCC) crystal structure. The crystal is in the cubic orientation with $X=[100], Y=[010], Z=[001]$ and the size of the model is $40a \times 8a \times 80a (14.05 \times 2.82 \times 28.16 nm)$, where $a (=3.52$ \AA) is lattice constant of nickel. As shown in Fig. \ref{fig:model}a, a crack is placed on the left edge of the model. The length of the initial crack is equal to $8a (2.82nm)$ and the width is $5a (1.76nm)$. Figure \ref{fig:model}b and \ref{fig:model}c are the atomic models of a mono-crystal with a void or inclusion respectively, where $R_v$ is the radius of the void and $R_i$ is the radius of the inclusion. In this study, in order to apply the mode \uppercase\expandafter{\romannumeral1} loading conditions, atoms in the top and bottom layers are fixed by freezing on the constant lattice sites and non-periodic boundary conditions are applied. The simulation process includes two steps as follows. Firstly, the initial atomic model was relaxed to minimize the energy. Secondly, a velocity-controlled tensile loading which is equivalent to a strain rate $r_\varepsilon$ was applied on the fixed atoms labeled in Fig. \ref{fig:model} and the simulation should be conducted at a constant temperature $t$, where the $r_\varepsilon$ and $t$ should be determined case by case. For comparison, a time-steps of 0.001 ps (picoseconds) is used in all the simulations. Moreover, the open source MD code, named Large-scale Atomic Molecular Massively Parallel Simulator (LAMMPS) \cite{plimpton1995fast}, is used to simulate the crack propagation of crystals and the atomic visualization of the MD simulation results is processed by an open source software termed Open Visualization Tool (OVITO) \cite{ovito}.

	\section{Results and discussions}
	
	\subsection{Edge crack propagation in Ni crystal} \label{section:edge}
	
\begin{figure*}[htbp]
		\centering
		\subfigure[$\varepsilon=5\%$]{\includegraphics[width=0.9in]{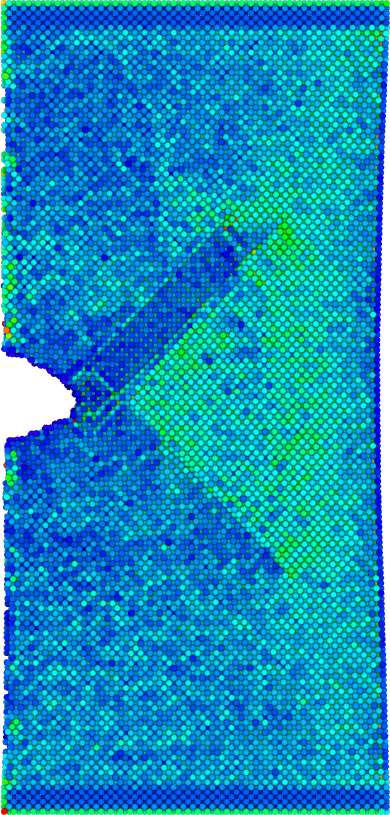}}
		\hspace{0.1in}
		\subfigure[$\varepsilon=10\%$]{\includegraphics[width=0.9in]{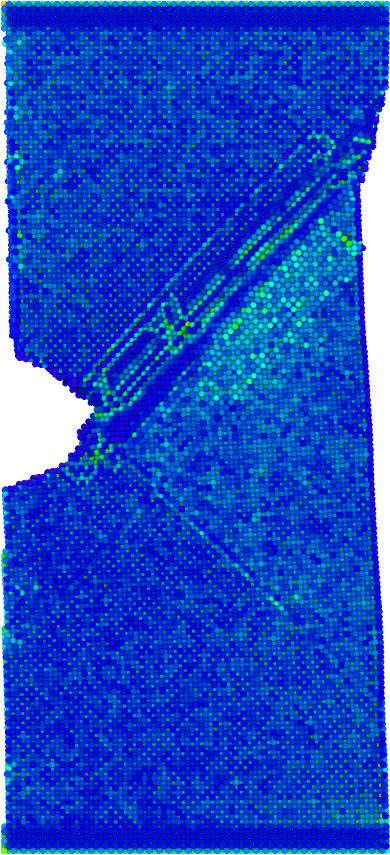}}
		\hspace{0.1in}
		\subfigure[$\varepsilon=13.33\%$]{\includegraphics[width=0.9in]{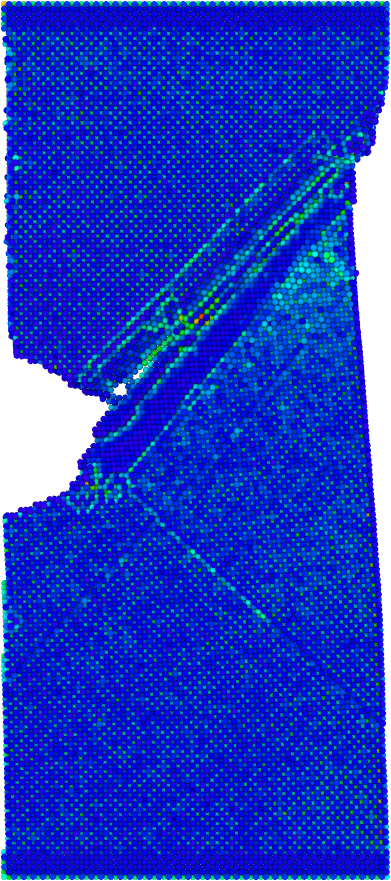}}
		\hspace{0.1in}
		\subfigure[$\varepsilon=16.67\%$]{\includegraphics[width=0.95in]{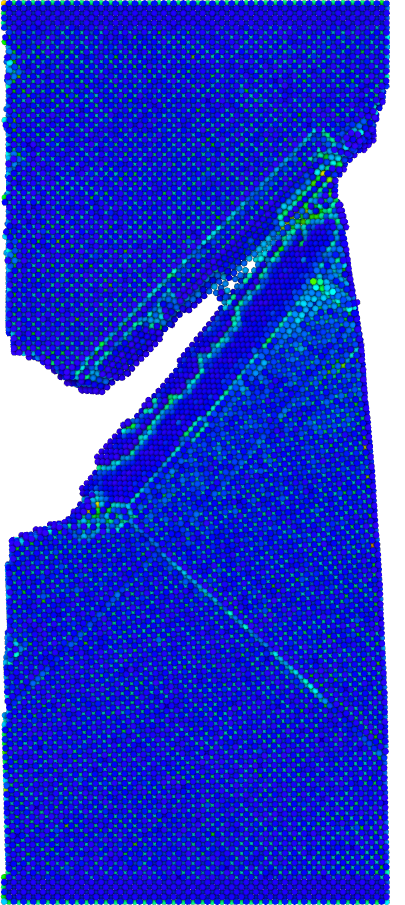}}
		\subfigure{\includegraphics[width=0.4in]{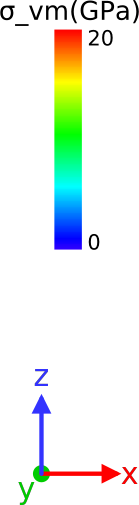}}
		\caption{Distributions of von Mises stress for strains of (a) 5\%, (b) 10\%, (c) 13.33\%, and (d) 16.67\% (strain rate $r_\varepsilon=6.67 \times 10^8s^{-1}$, temperature $t=50K$).}
		\label{fig:case1-stress}
	\end{figure*}

\begin{figure*}
		\centering
		\subfigure[Displacement contour]{\includegraphics[width=2in]{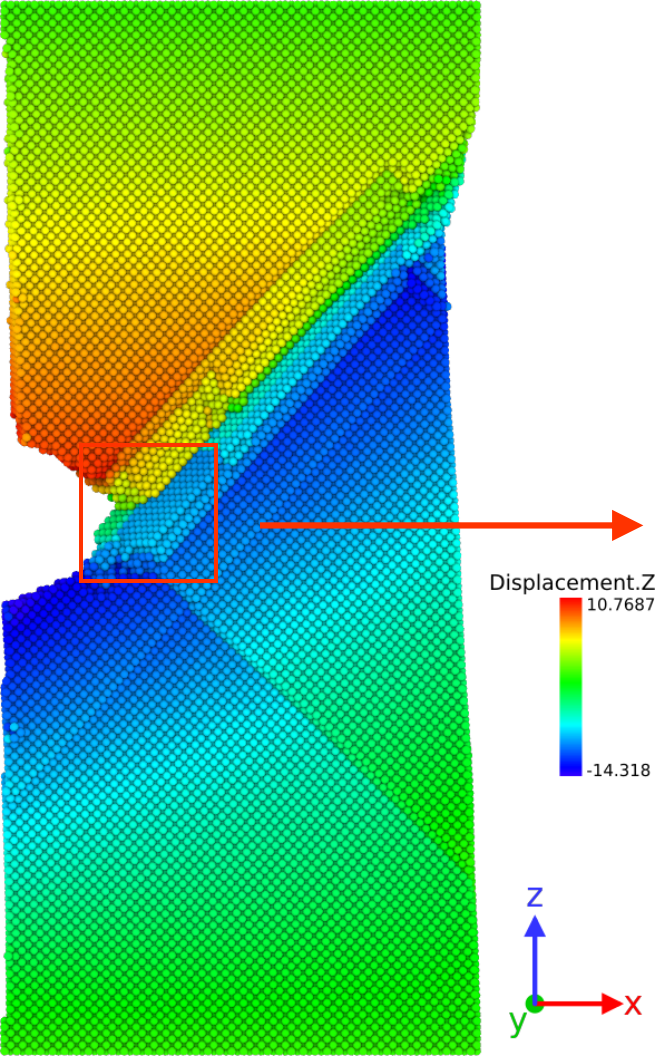}}
		\subfigure[Atomic displacement vector]{\includegraphics[width=2in]{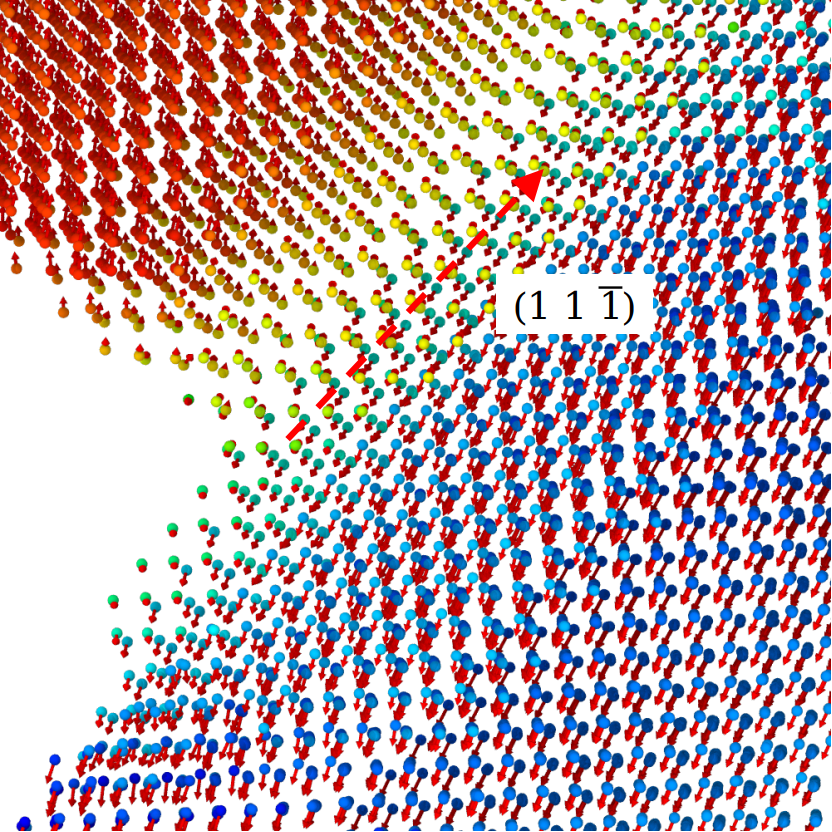}}
		\caption{Distributions of displacement in z direction for strain of 10\%, where (a) is the displacement contour and (b) is the atomic displacement vector around the crack tip.}
		\label{fig:dispz}
	\end{figure*}

\begin{figure*}[htbp]
		\centering
		\includegraphics[width=3in]{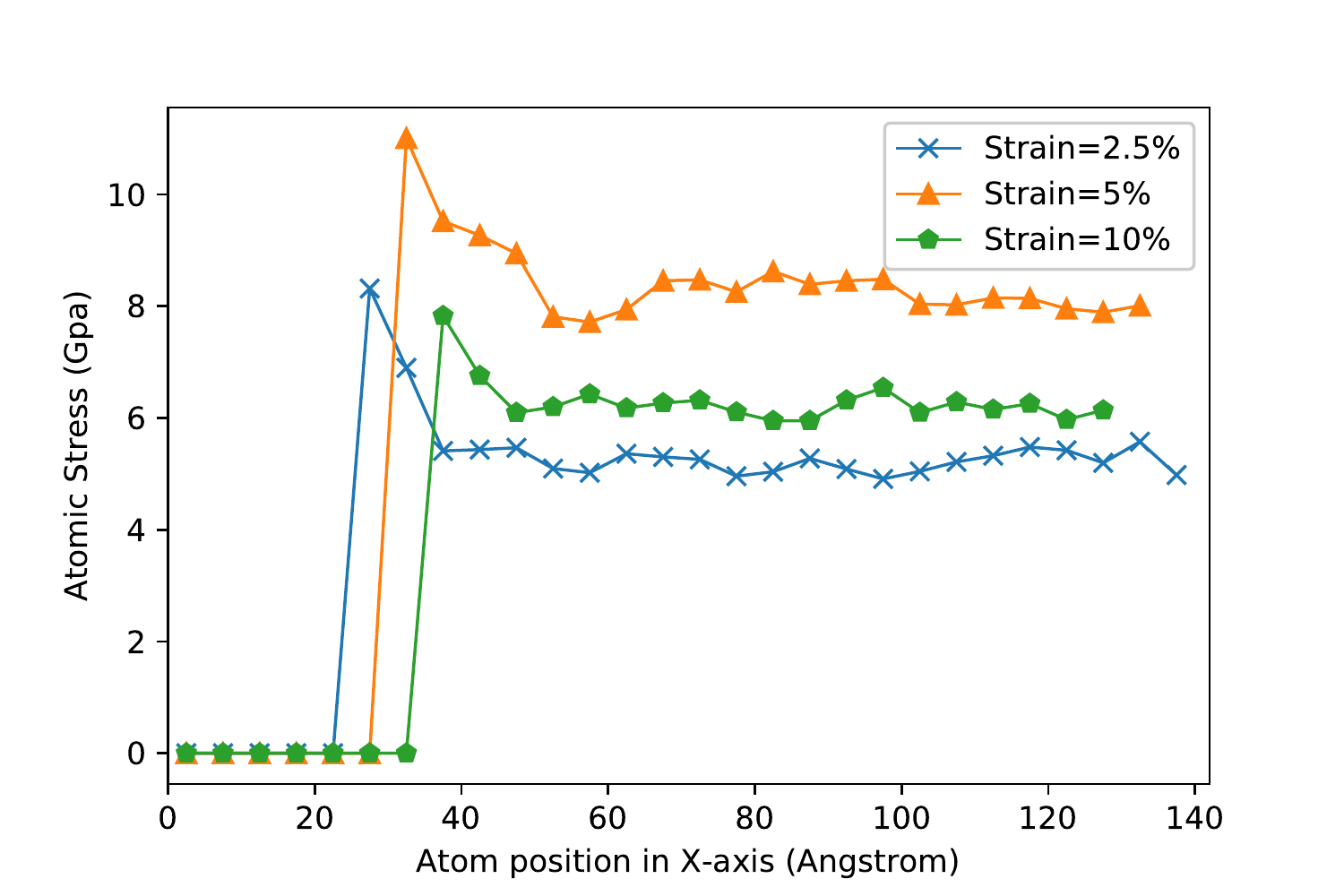}
		\caption{Atomic stress as function of the atom position in X-axis for crack propagation in a Ni crystal.}
		\label{fig:cts_ed}
	\end{figure*}

\begin{figure*}[htb]
		\centering
		\includegraphics[width=4.5in]{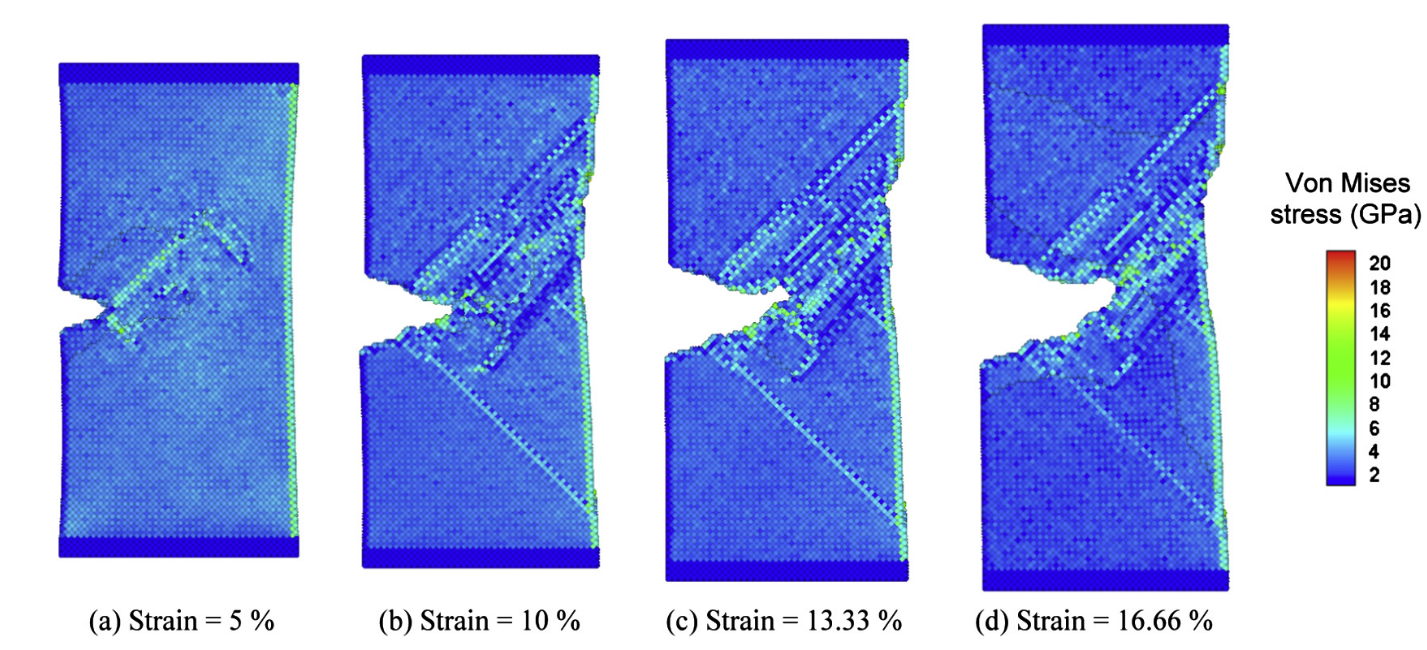}
		\caption{Simulation result of reference \cite{sung2015studies} with strain rate $r_\varepsilon=6.67 \times 10^8s^{-1}$, temperature $t=50K$.}
		\label{fig:case1-ref}
	\end{figure*}

	As shown in Fig. \ref{fig:model}, a mono-crystal of Ni with an edge crack is considered in this study. In order to make a comparison with the results of the reference \cite{sung2015studies}, the constant strain rate $r_\varepsilon$ of $6.67 \times 10^8s^{-1}$ and the temperature $t=50K$ are used in this case. Figure \ref{fig:case1-stress} shows the von Mises stress distributions of the edge crack propagation under various strains. The atoms are colored by the magnitude of von Mises stress. It can be clearly seen that the stress concentration forms at the crack tip and dislocations grow from the crack tip when the strain is 5\% as shown in Fig. \ref{fig:case1-stress}a. Figure \ref{fig:case1-stress}b shows that the crack significantly advances into the close-paced plane ($1 1 \bar{1}$) for the FCC structure. The details are shown as Fig. \ref{fig:dispz}, where the atoms are colored by the magnitude of displacement in z direction and Fig. \ref{fig:dispz}b shows the atomic displacement vector around the crack tip. It is obvious that the atoms slip along the ($1 1 \bar{1}$) plane, which matches the results of reference \cite{sung2015studies}. As for Fig. \ref{fig:dispz}c, it shows that the crack starts growing along the slipping plane when the strain is increased to 13.33\%, which means that the stress reaches the critical stress. Figure \ref{fig:cts_ed} presents a quantitative analysis of atomic stress around the crack tip. The atomic stress is plotted as functions of the atom position along the X-axis. It can be clearly seen that the atomic stress concentrated directly at the crack tip and the crack tip moves with the increment of the strain. Compared with the results of reference \cite{sung2015studies}, which is shown in Fig. \ref{fig:case1-ref}, the results match with each other, especially in slip bands and dislocation locations.
	
	\subsection{Edge crack propagation in Ni crystal with voids}
	
\begin{figure*}[htbp]
		\centering
		\subfigure[$\varepsilon=5\%$]{\includegraphics[width=0.9in]{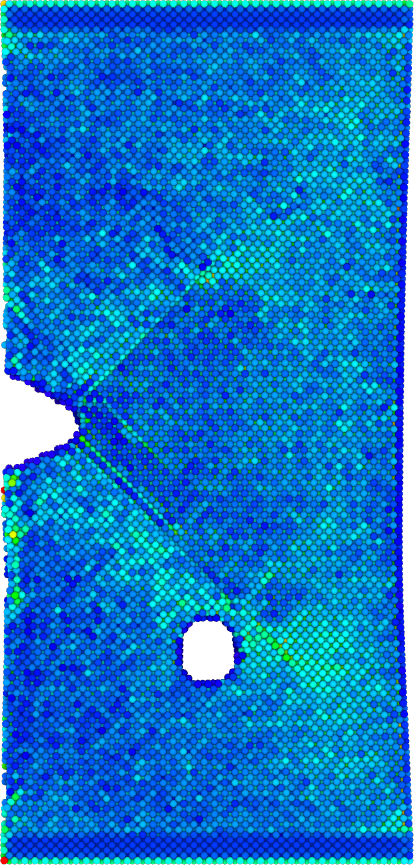}}
		\hspace{0.1in}
		\subfigure[$\varepsilon=10\%$]{\includegraphics[width=0.9in]{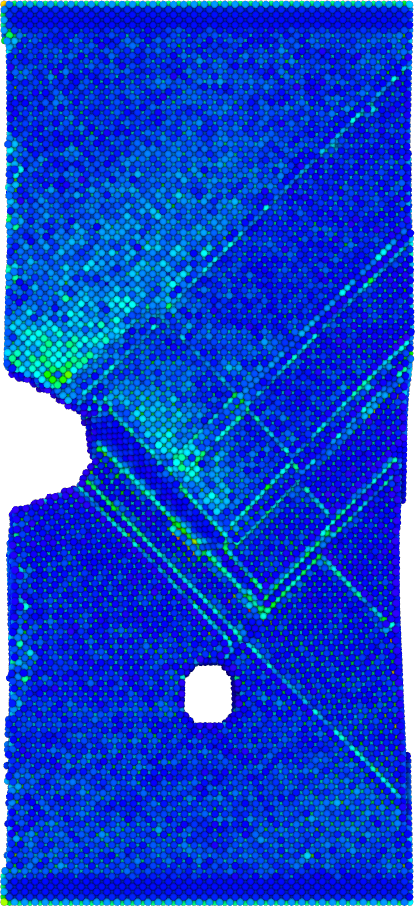}}
		\hspace{0.1in}
		\subfigure[$\varepsilon=13.33\%$]{\includegraphics[width=0.9in]{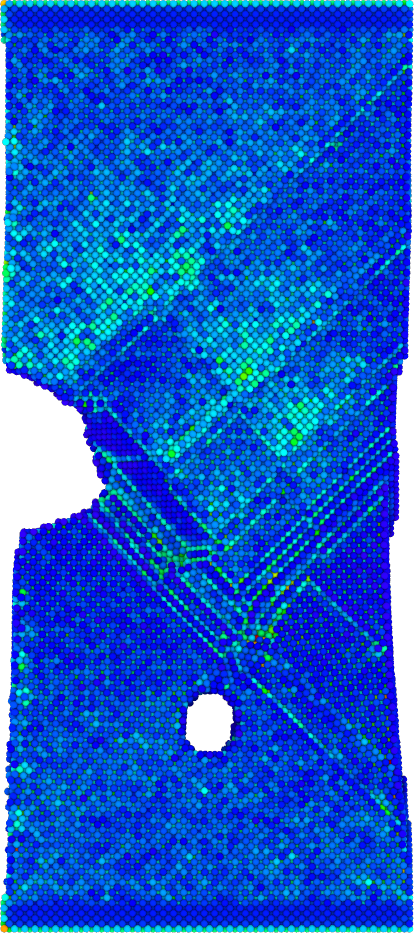}}
		\hspace{0.1in}
		\subfigure[$\varepsilon=16.67\%$]{\includegraphics[width=0.9in]{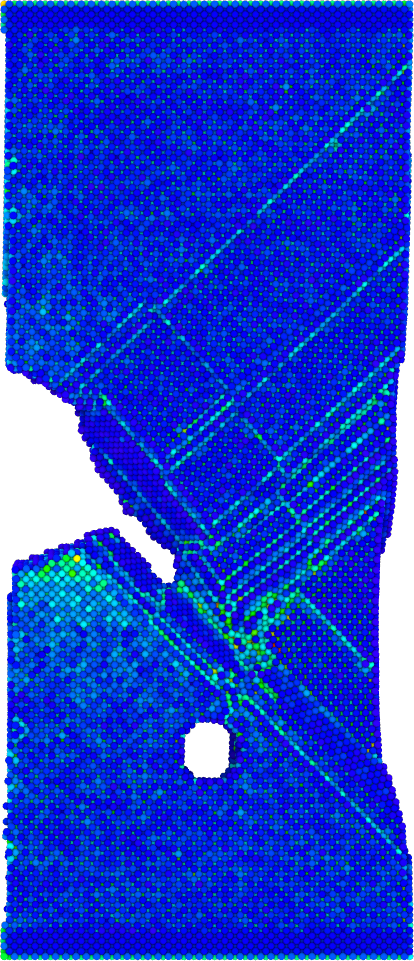}}
		\subfigure{\includegraphics[width=0.4in]{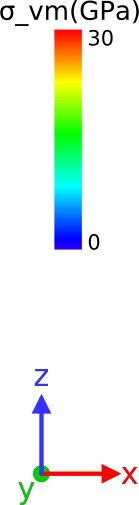}}
		\caption{Distributions of von Mises stress for crack propagation in a mono-crystal with a void of radius 3 \AA: (a) 5\%, (b) 10\%, (c) 13.33\%, and (d) 16.67\% (strain rate $r_\varepsilon=6.67 \times 10^8s^{-1}$, temperature $t=50K$).}
		\label{fig:case2-vr3-stress}
	\end{figure*}

\begin{figure*}[htbp]
		\centering	
		\subfigure[$\varepsilon=5\%$]{\includegraphics[width=0.9in]{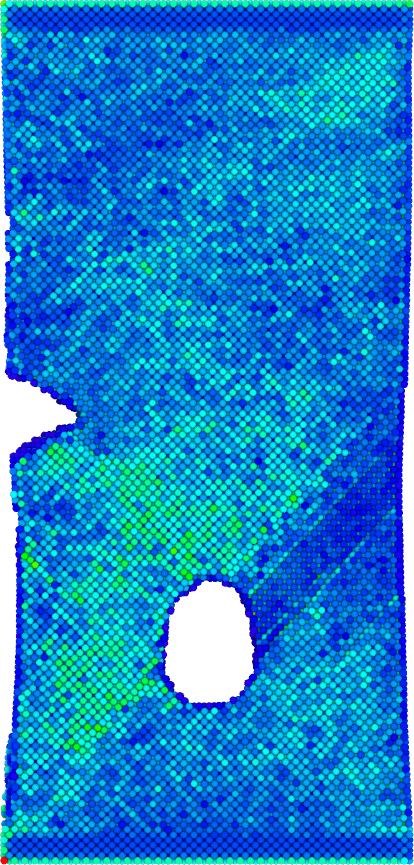}}
		\hspace{0.1in}
		\subfigure[$\varepsilon=10\%$]{\includegraphics[width=0.9in]{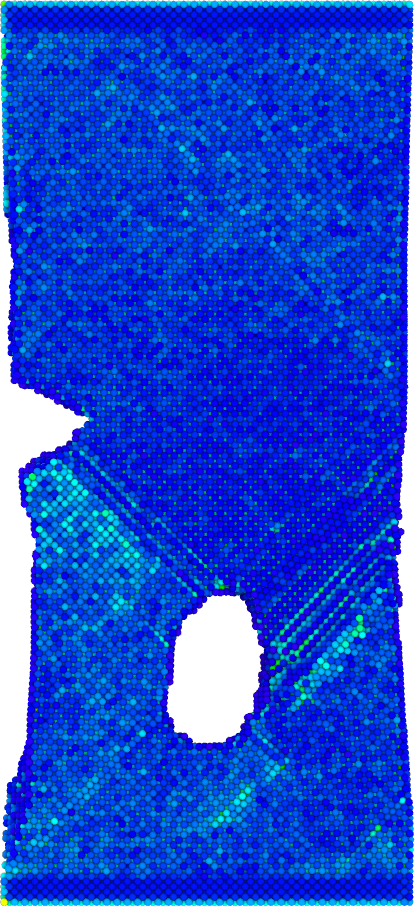}}
		\hspace{0.1in}
		\subfigure[$\varepsilon=13.33\%$]{\includegraphics[width=0.9in]{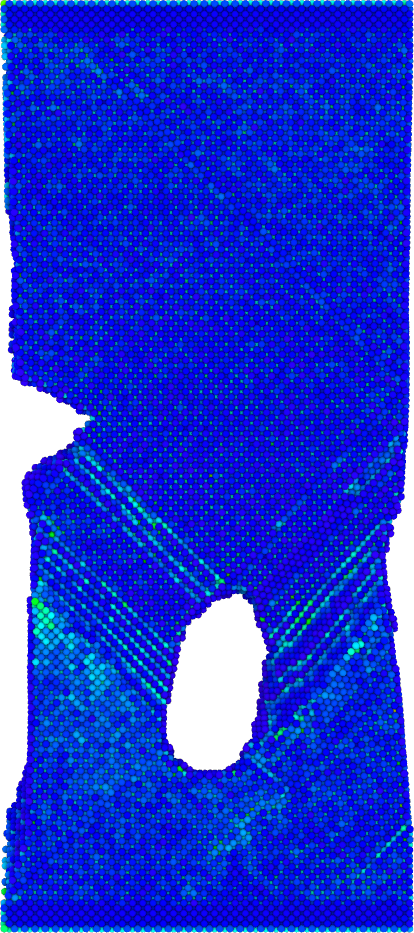}}
		\hspace{0.1in}
		\subfigure[$\varepsilon=16.67\%$]{\includegraphics[width=0.9in]{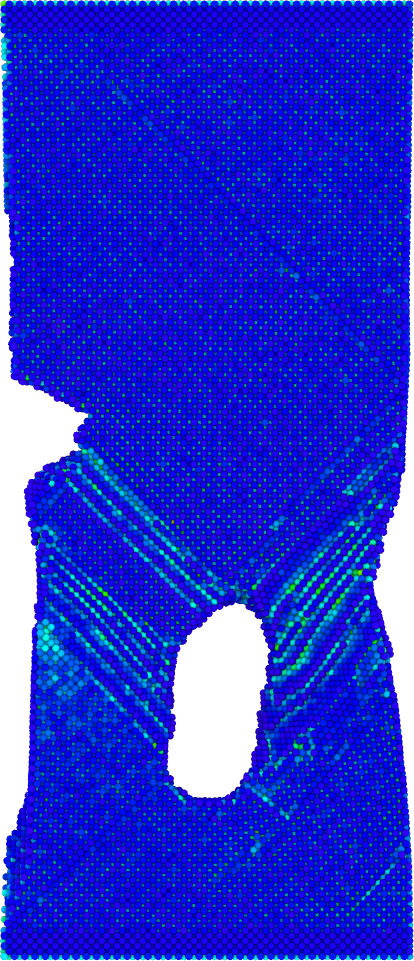}}
		\subfigure{\includegraphics[width=0.4in]{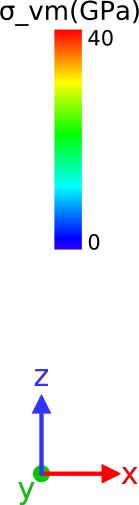}}\\
		\caption{Distributions of von Mises stress for crack propagation in a mono-crystal with a void of radius 5 \AA: (a) 5\%, (b) 10\%, (c) 13.33\%, and (d) 16.67\% (strain rate $r_\varepsilon=6.67 \times 10^8s^{-1}$, temperature $t=50K$).}
		\label{fig:case2-vr5-stress}
	\end{figure*}

\begin{figure*}[htbp]
		\centering
		\subfigure[$\varepsilon=5\%$]{\includegraphics[width=0.9in]{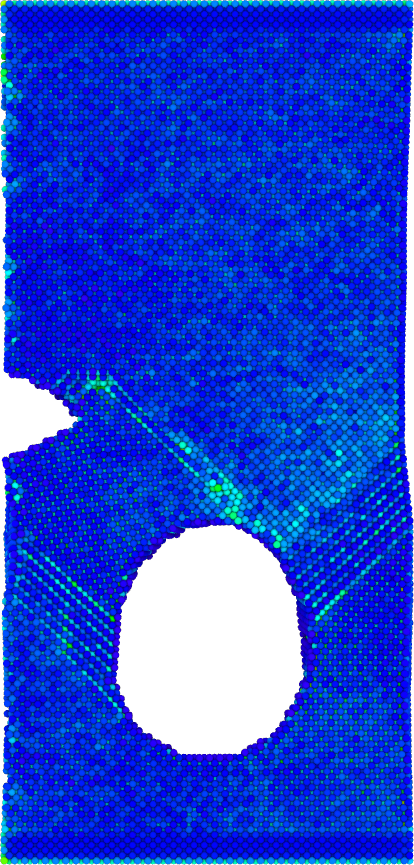}}
		\hspace{0.1in}
		\subfigure[$\varepsilon=10\%$]{\includegraphics[width=0.9in]{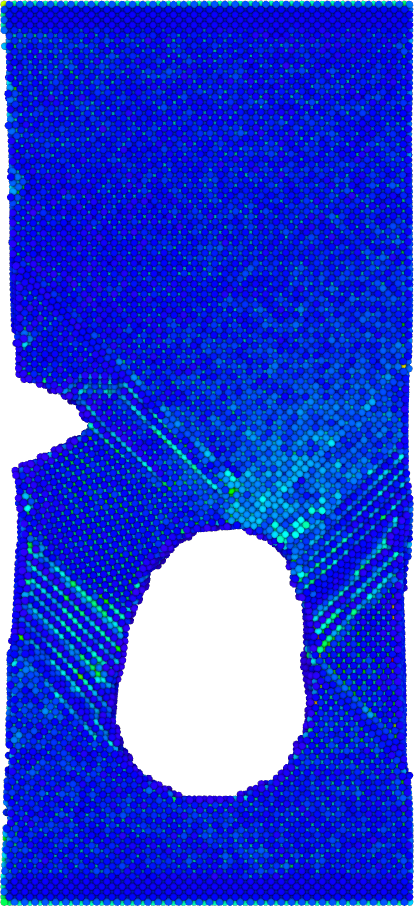}}
		\hspace{0.1in}
		\subfigure[$\varepsilon=13.33\%$]{\includegraphics[width=0.9in]{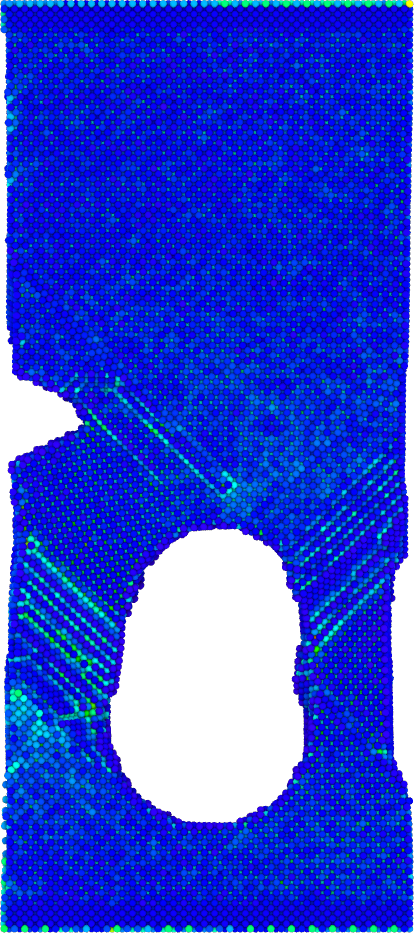}}
		\hspace{0.1in}
		\subfigure[$\varepsilon=16.67\%$]{\includegraphics[width=0.9in]{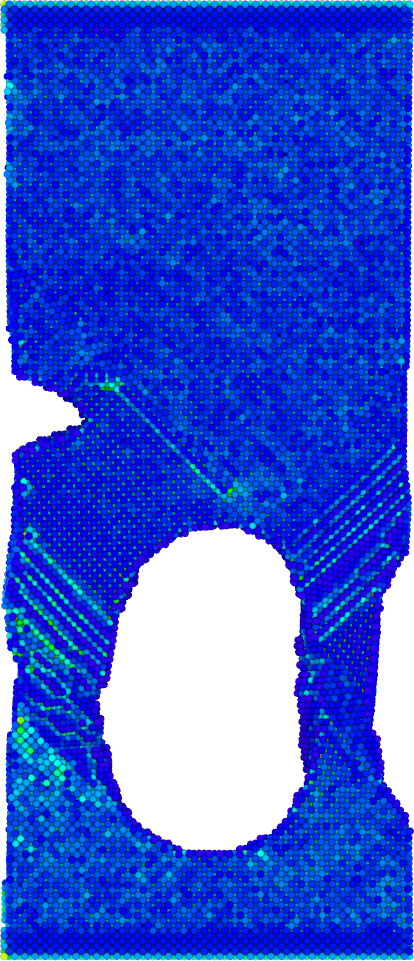}}
		\subfigure{\includegraphics[width=0.4in]{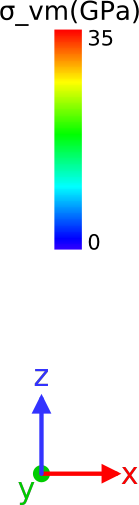}}
		\caption{Distributions of von Mises stress for crack propagation in a mono-crystal with a void of radius 10 \AA: (a) 5\%, (b) 10\%, (c) 13.33\%, and (d) 16.67\% (strain rate $r_\varepsilon=6.67 \times 10^8s^{-1}$, temperature $t=50K$).}
		\label{fig:case2-vr10-stress}
	\end{figure*}

\begin{figure*}[htbp]
		\centering
		\includegraphics[width=3in]{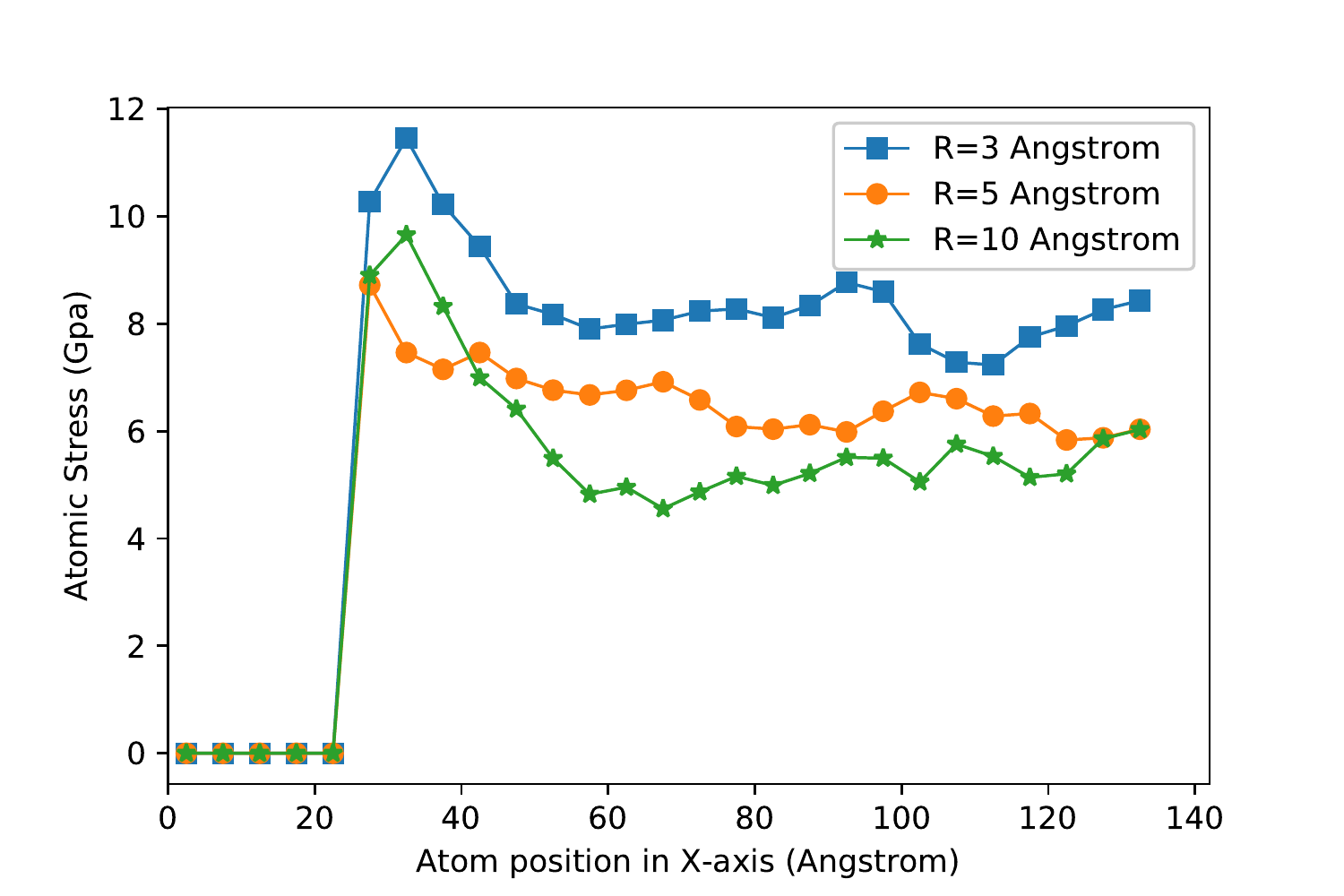}
		\caption{Atomic stress as function of the atom position in X-axis at strain of 5\% for crack propagation in a Ni crystal with voids.}
		\label{fig:cts_voids}
	\end{figure*}

	To study the effect of voids on the mechanical properties of crack propagation for a mono-crystal, an atomic model of a crystal with a void is built as shown in Fig. \ref{fig:model}b and different radii of voids are used to fully investigate the effect of voids. The strain rate of $6.67 \times 10^8s^{-1}$ and temperature of $50K$ are also used to make a comparison with the results in section \ref{section:edge}. Figure \ref{fig:case2-vr3-stress} shows the distributions of von Mises stress for crack propagation in a crystal with a void of $R_v=3$\AA. Obviously, the stress concentration and dislocations develop at the strain of 5\% and then the dislocations grow and emit along the close-packed plane when the strain is increased to 10\%. Figure \ref{fig:case2-vr3-stress}c shows that the crack is just about to grow and Fig. \ref{fig:case2-vr3-stress}d shows that the crack grows along the closed-packed plane. More importantly, it can be determined that the crack advances into another direction due to the effect of the void. Compare with the results in section \ref{section:edge}, the crack grows along the $(\bar{1} 1 \bar{1})$ plane, no longer the $(1 1 \bar{1})$ plane. As shown in Fig. \ref{fig:case2-vr5-stress}, a void of $R_v=5$\AA is used in this case. It can be determined that there are no significant dislocations around the crack tip due to the effect of the void at the strain of 5\%. Figure \ref{fig:case2-vr5-stress}b and \ref{fig:case2-vr5-stress}c show that dislocations develop from the crack tip and the void when the strain is increased. Moreover, there is no significant phenomenon of the crack propagation when the strain is increased up to 16.67\%, but it can be clearly seen that the void is deformed. As for Fig. \ref{fig:case2-vr10-stress}, it shows the distributions of von Mises stress for crack propagation in a crystal with a void of $R_v=10$\AA. It can be found that the similar phenomenon is observed when comparing with Fig. \ref{fig:case2-vr5-stress}. There are no significant dislocations around the crack tip but some dislocations grow around the void and no significant crack propagation has been observed even the strain is up to 16.67\%. Figure \ref{fig:cts_voids} shows the atomic stress as a function of the atom position in X-axis at the strain of 5\%. It is obvious that the atomic stress concentrated at the crack tip as described above. Therefore, the results indicate that voids can change the path of crack propagation. For small voids, it can be used to guide the path of crack propagation. For some larger voids, the results show that the voids can absorb the strain energy, which can lead a better resistance to plastic deformation in crystals.
	
	\subsection{Edge crack propagation in Ni crystal with inclusions}
	
\begin{figure*}[htbp]
		\centering
		\subfigure[$R_i=3$\AA, $\varepsilon=10\%$]{\includegraphics[width=0.9in]{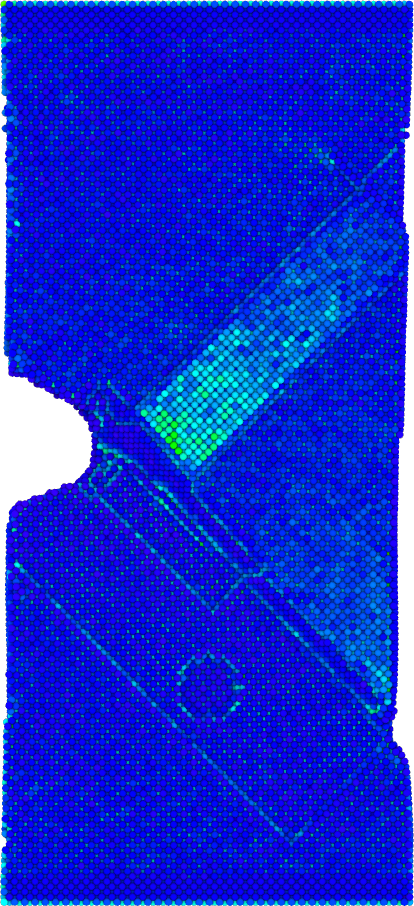}}
		\hspace{0.4in}
		\subfigure[$R_i=5$\AA, $\varepsilon=10\%$]{\includegraphics[width=0.9in]{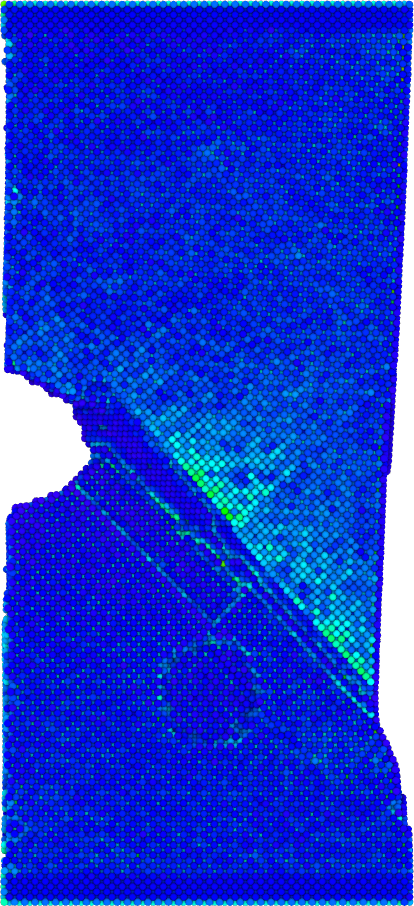}}
		\hspace{0.4in}
		\subfigure[$R_i=10$\AA, $\varepsilon=10\%$]{\includegraphics[width=0.9in]{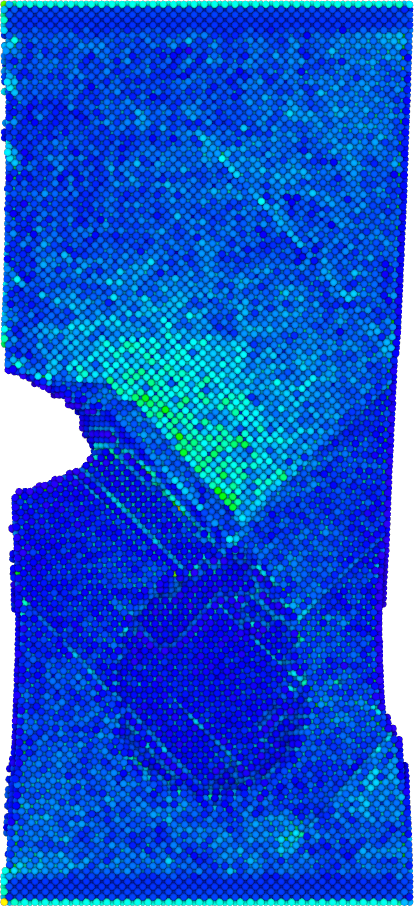}}
		\subfigure{\includegraphics[width=0.4in]{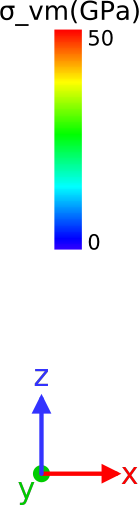}} \\
		
		\subfigure[$R_i=3$\AA, $\varepsilon=16.67\%$]{\includegraphics[width=0.9in]{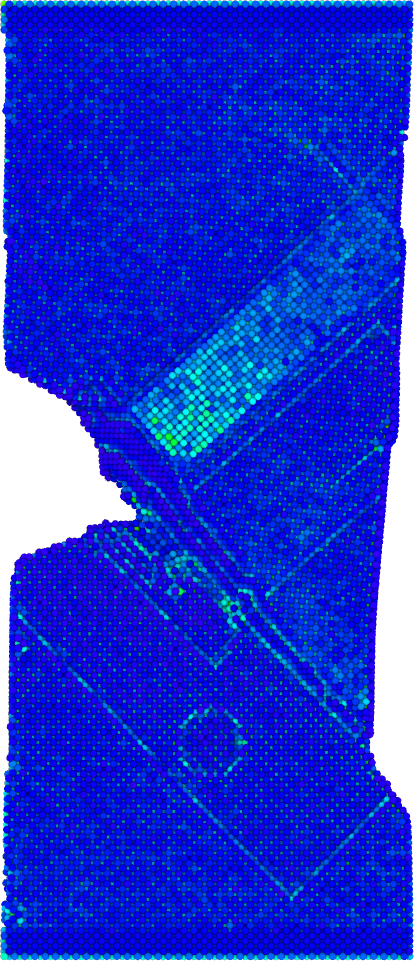}}
		\hspace{0.4in}
		\subfigure[$R_i=5$\AA, $\varepsilon=16.67\%$]{\includegraphics[width=0.9in]{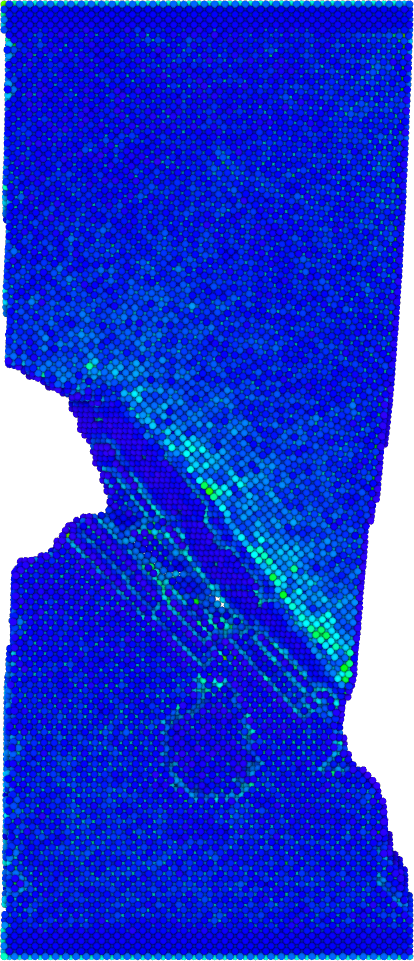}}
		\hspace{0.4in}
		\subfigure[$R_i=10$\AA, $\varepsilon=16.67\%$]{\includegraphics[width=0.9in]{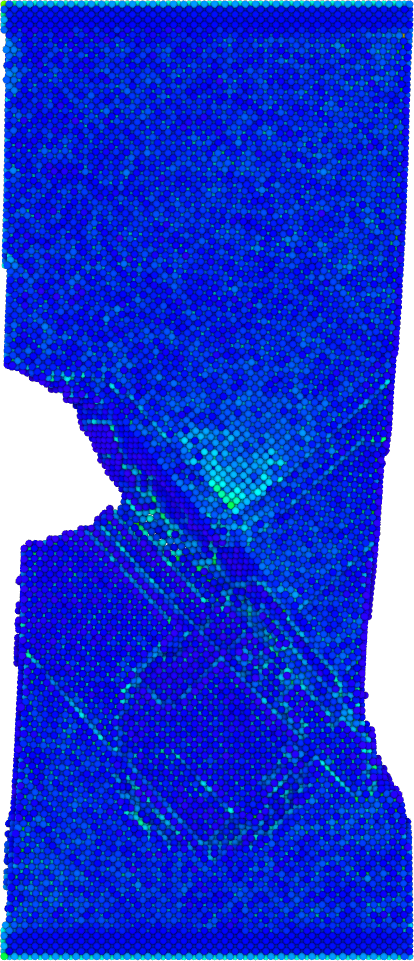}}
		\subfigure{\includegraphics[width=0.4in]{case3-ir3-legend.png}}
		\caption{Distributions of von Mises stress for crack propagation in a mono-crystal with copper inclusions (strain rate $r_\varepsilon=6.67 \times 10^8s^{-1}$, temperature $t=50K$).}
		\label{fig:case3-cu-stress}
	\end{figure*}

\begin{figure*}[htbp]
		\centering
		\subfigure[$R_i=3$\AA, $\varepsilon=10\%$]{\includegraphics[width=0.9in]{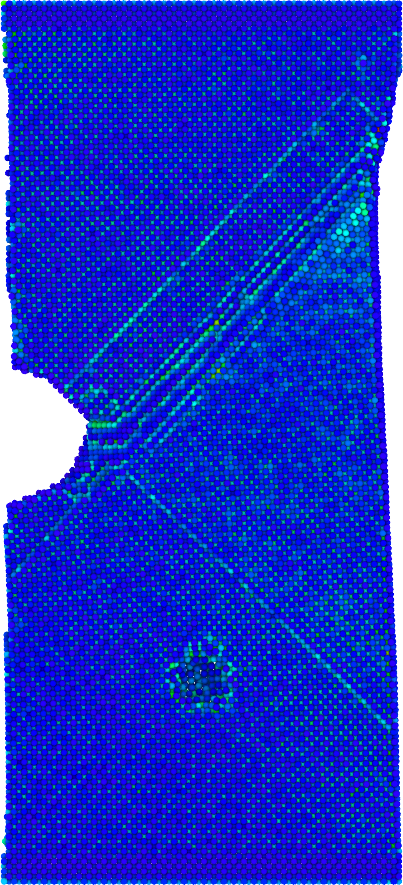}}
		\hspace{0.4in}
		\subfigure[$R_i=5$\AA, $\varepsilon=10\%$]{\includegraphics[width=0.9in]{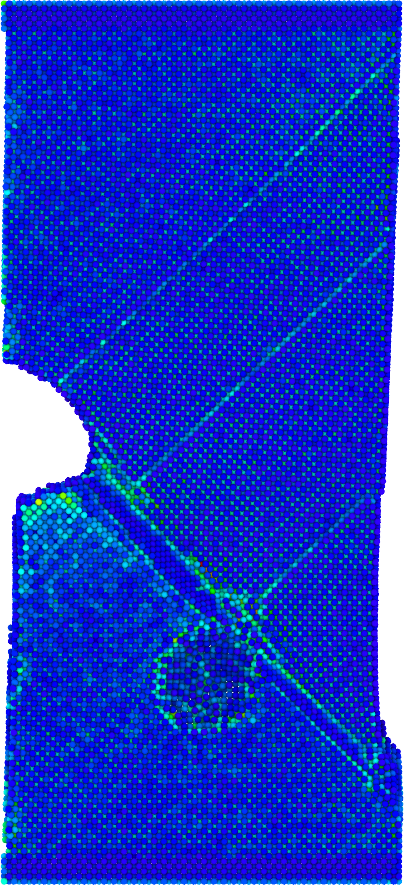}}
		\hspace{0.4in}
		\subfigure[$R_i=10$\AA, $\varepsilon=10\%$]{\includegraphics[width=0.9in]{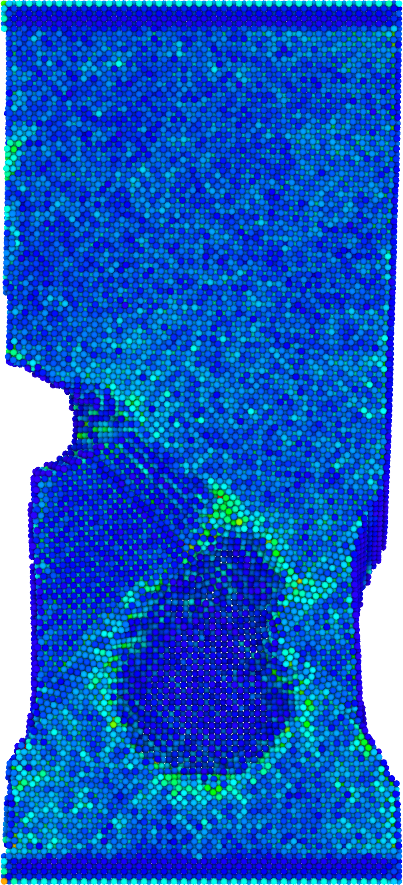}}
		\subfigure{\includegraphics[width=0.4in]{case3-ir3-legend.png}} \\
		
		\subfigure[$R_i=3$\AA, $\varepsilon=16.67\%$]{\includegraphics[width=0.9in]{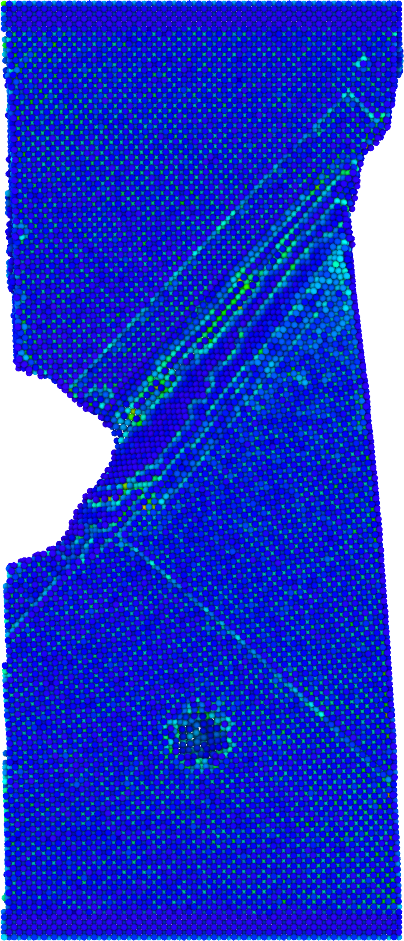}}
		\hspace{0.4in}
		\subfigure[$R_i=5$\AA, $\varepsilon=16.67\%$]{\includegraphics[width=0.9in]{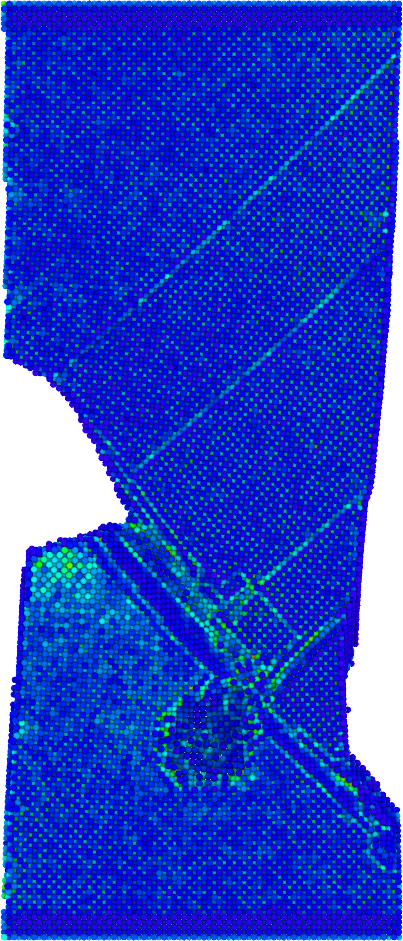}}
		\hspace{0.4in}
		\subfigure[$R_i=10$\AA, $\varepsilon=16.67\%$]{\includegraphics[width=0.9in]{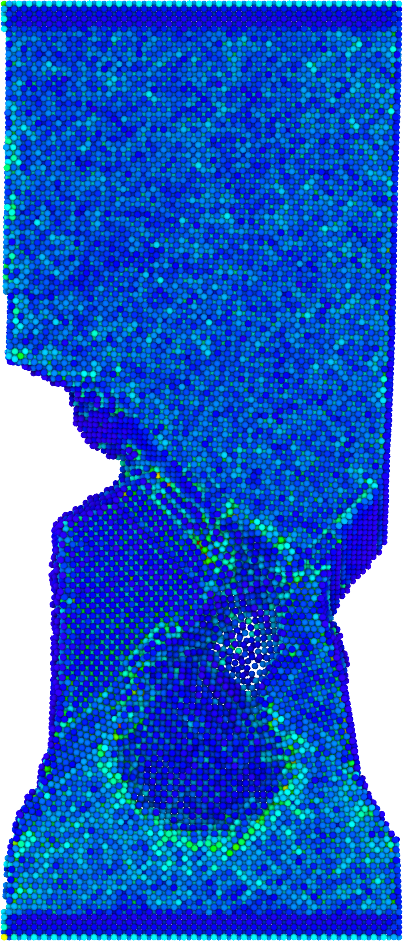}}
		\subfigure{\includegraphics[width=0.4in]{case2-vr5-legend.png}}
		\caption{Distributions of von Mises stress for crack propagation in a mono-crystal with aluminum inclusions (strain rate $r_\varepsilon=6.67 \times 10^8s^{-1}$, temperature $t=50K$).}
		\label{fig:case3-al-stress}
	\end{figure*}

\begin{figure*}[htbp]
		\centering
		\subfigure[Cu inclusions]{\includegraphics[width=2.3in]{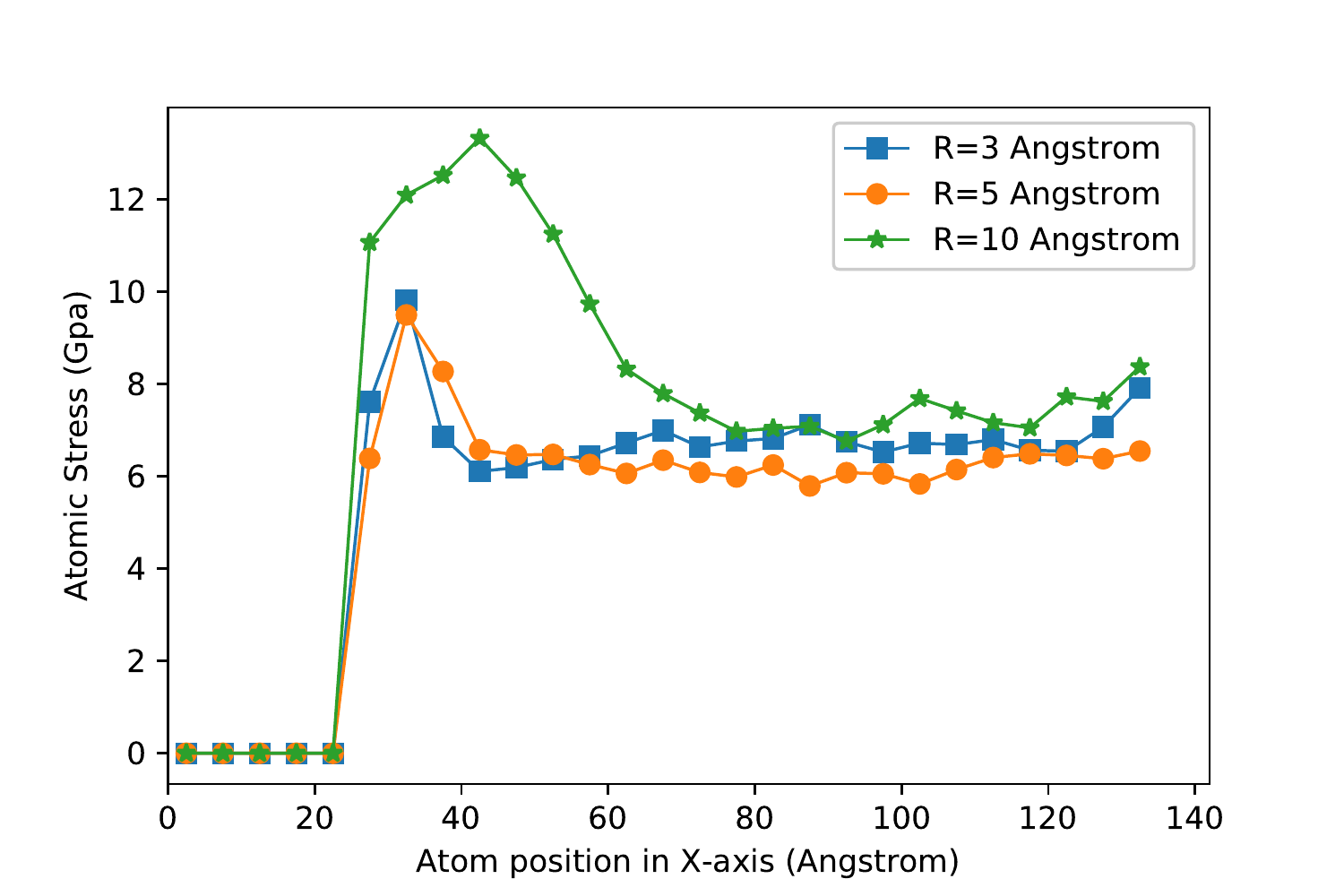}}
		\subfigure[Al inclusions]{\includegraphics[width=2.3in]{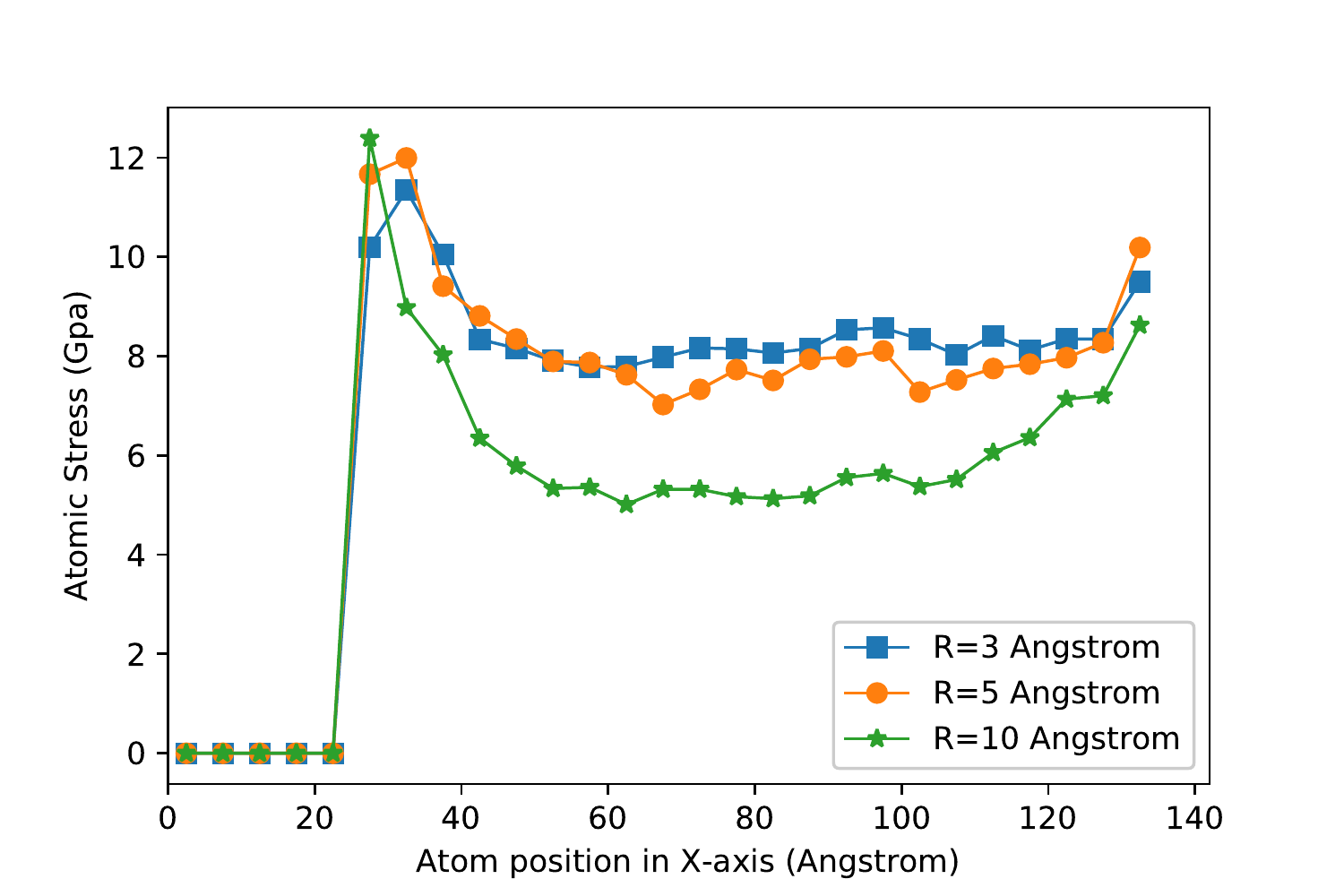}}
		\caption{Atomic stress as function of the atom position in X-axis at strain of 5\% for crack propagation in a Ni crystal with (a) copper inclusions and (b) aluminum inclusions.}
		\label{fig:cts_inc}
	\end{figure*}

	In order to investigate the effect of inclusions on the edge crack propagation in crystals, an atomic model of crystal with a cylindrical inclusion is built as shown Fig. \ref{fig:model}c. Moreover, different radii and materials of cylindrical inclusions are used to fully study the effect of inclusions. Two metal materials, copper and aluminum are used in this case. Copper and aluminum are both metal materials with FCC crystal structure and the lattice constants are 3.61 \AA and 4.05 \AA respectively. In order to make a comparison with the results proposed above, the stain rate of $6.67 \times 10^8 s^{-1}$ and the temperature of $50K$ are used in this case. Figure \ref{fig:case3-cu-stress} shows the distributions of von Mises stress for crack propagation in a Ni crystal with the Cu inclusion, where $R_i$ means the radius of cylindrical inclusion and $\varepsilon$ is the strain. It can be determined that the stress concentration forms around the crack tip as shown in Fig \ref{fig:cts_inc}. More importantly, it can be clearly seen that the direction of crack propagation was changed to along the ($\bar{1} 1 \bar{1}$) plane and the dislocations were growing around the inclusions. Compared with Fig. \ref{fig:case1-stress}, the crack just gets a small increment at the strain of 16.67\% while the crack extends in a long distance in Fig. \ref{fig:case1-stress}. Therefore, the results show that the inclusions can lead a better resistance to plastic deformation during the crack propagation process. As for Fig. \ref{fig:case3-al-stress}, it shows the distributions of von Mises stress for crack propagation in a Ni crystal with the Al inclusion. It can be found that the direction of crack propagation was changed to along the ($\bar{1} 1 \bar{1}$) plane except for the model with the inclusion of $R_i=3$\AA. Compared with Fig. \ref{fig:case1-stress}, it is obvious that the inclusions can lead a better resistance to plastic deformation during the crack propagation process since the crack just gets a small increment at the strain of 16.67\% due to the effect of inclusions.

	\subsection{Discussions about the results}
	
\begin{figure*}[htbp]
		\centering
		\subfigure[without voids and inclusions]{\includegraphics[width=2.3in]{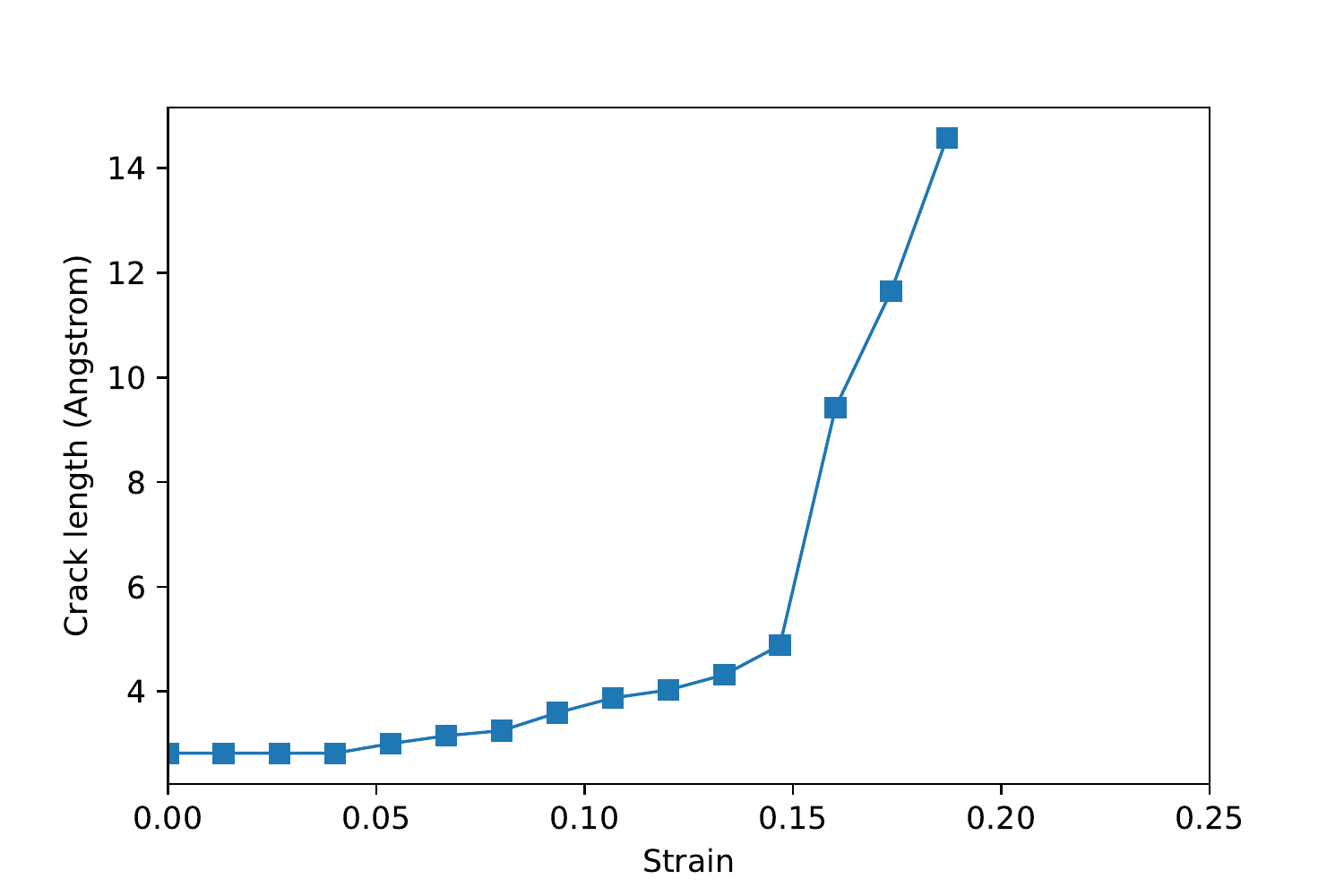}}
		\subfigure[with voids]{\includegraphics[width=2.3in]{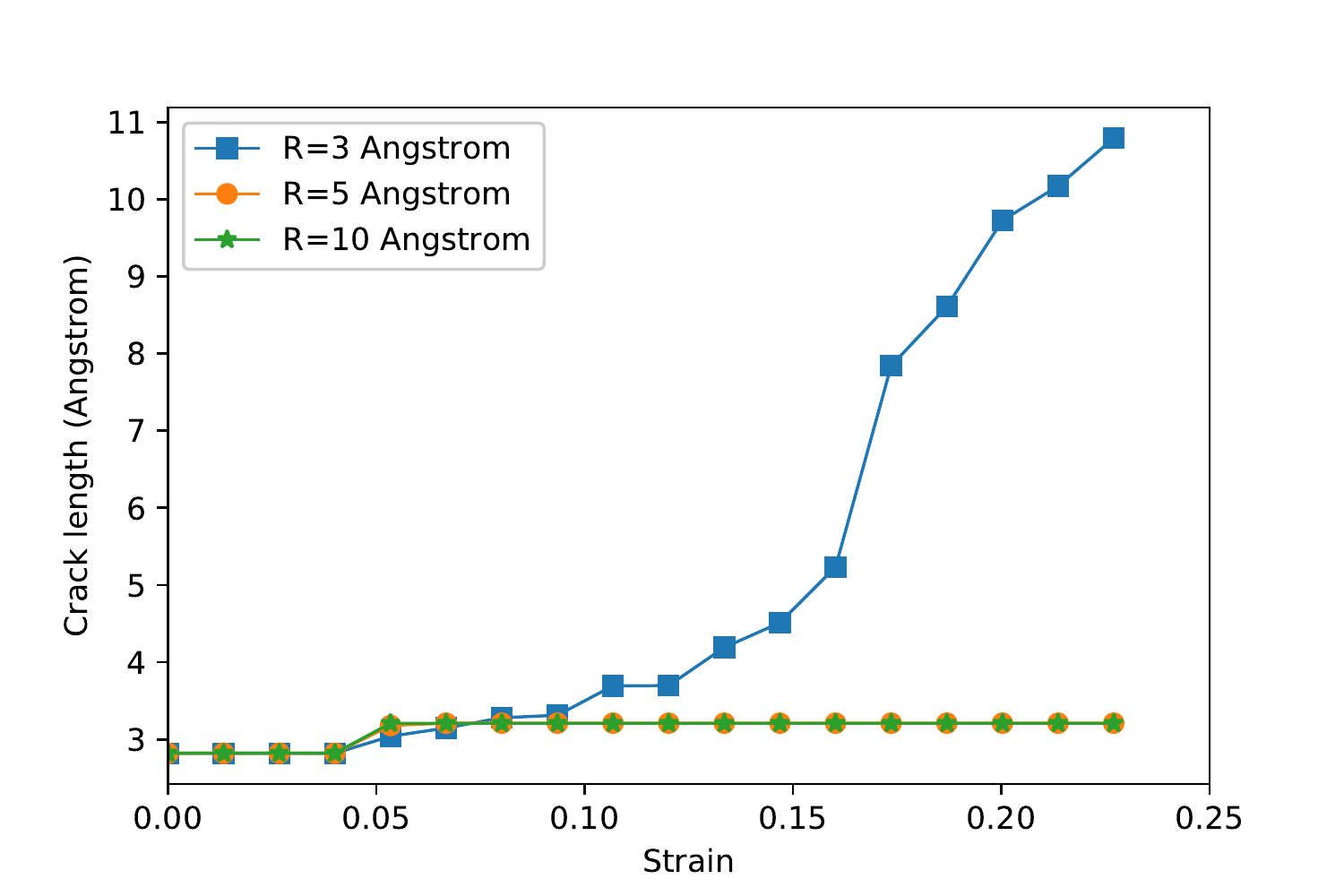}}\\
		
		\subfigure[with Cu inclusions]{\includegraphics[width=2.3in]{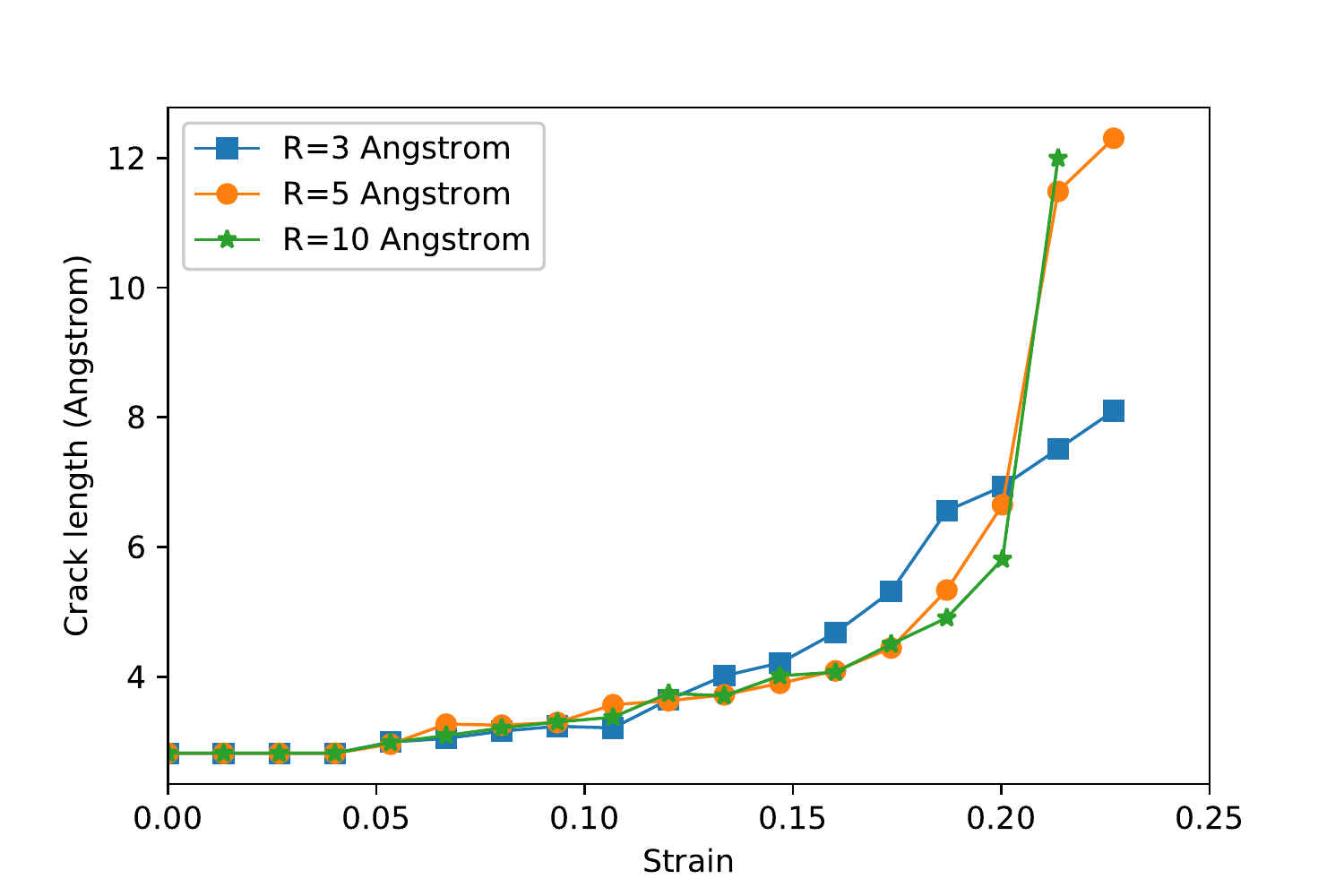}}
		\subfigure[with Al inclusions]{\includegraphics[width=2.3in]{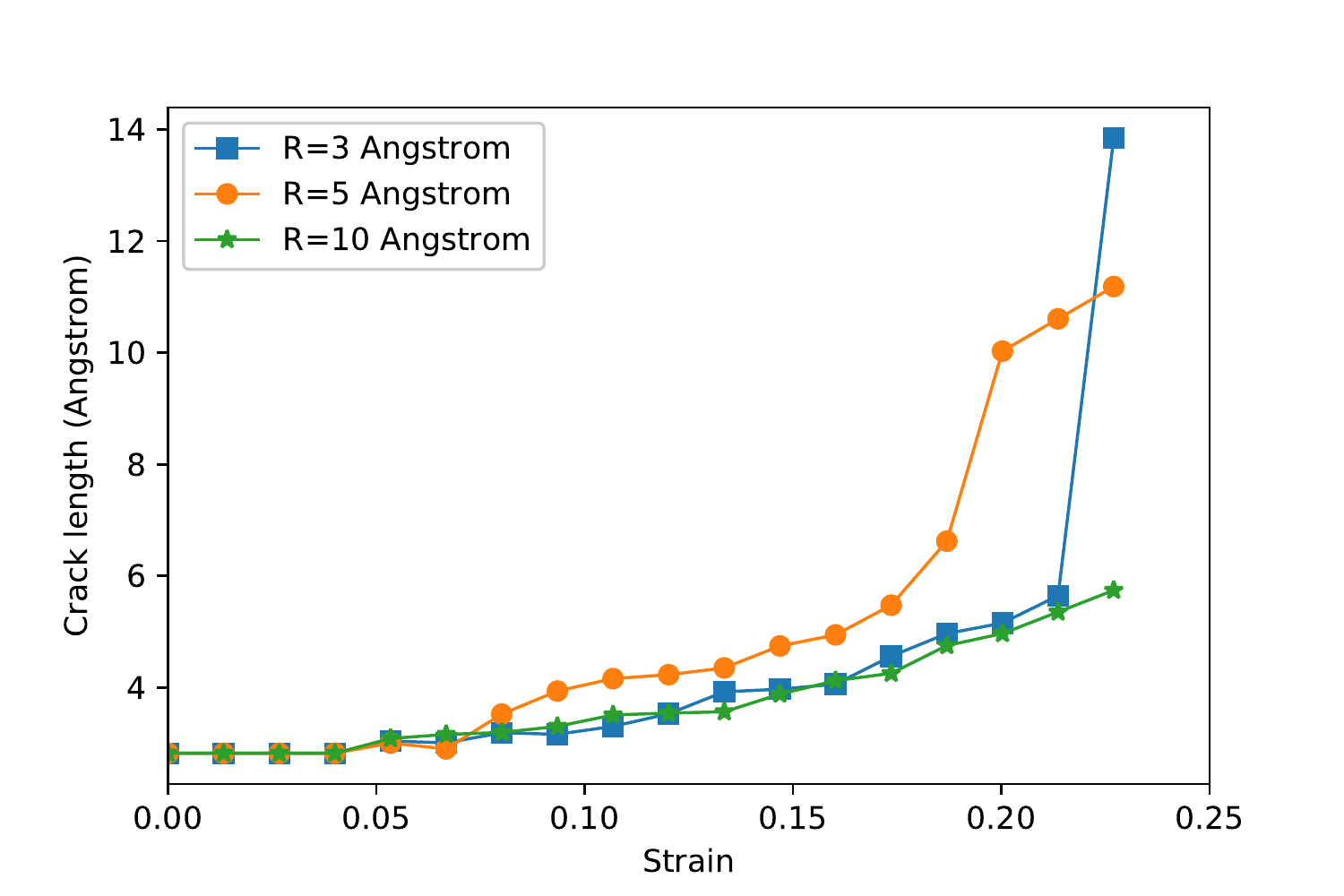}}
		\caption{Crack length as a function of strain for crack growing in a single crystal of nickel (a) without voids and inclusions, (b) with voids, (c) with Cu inclusions and (d) with Al inclusions.}
		\label{fig:crack_l}
	\end{figure*}

\begin{figure*}[htbp]
		\centering
		\subfigure[without voids and inclusions]{\includegraphics[width=2.3in]{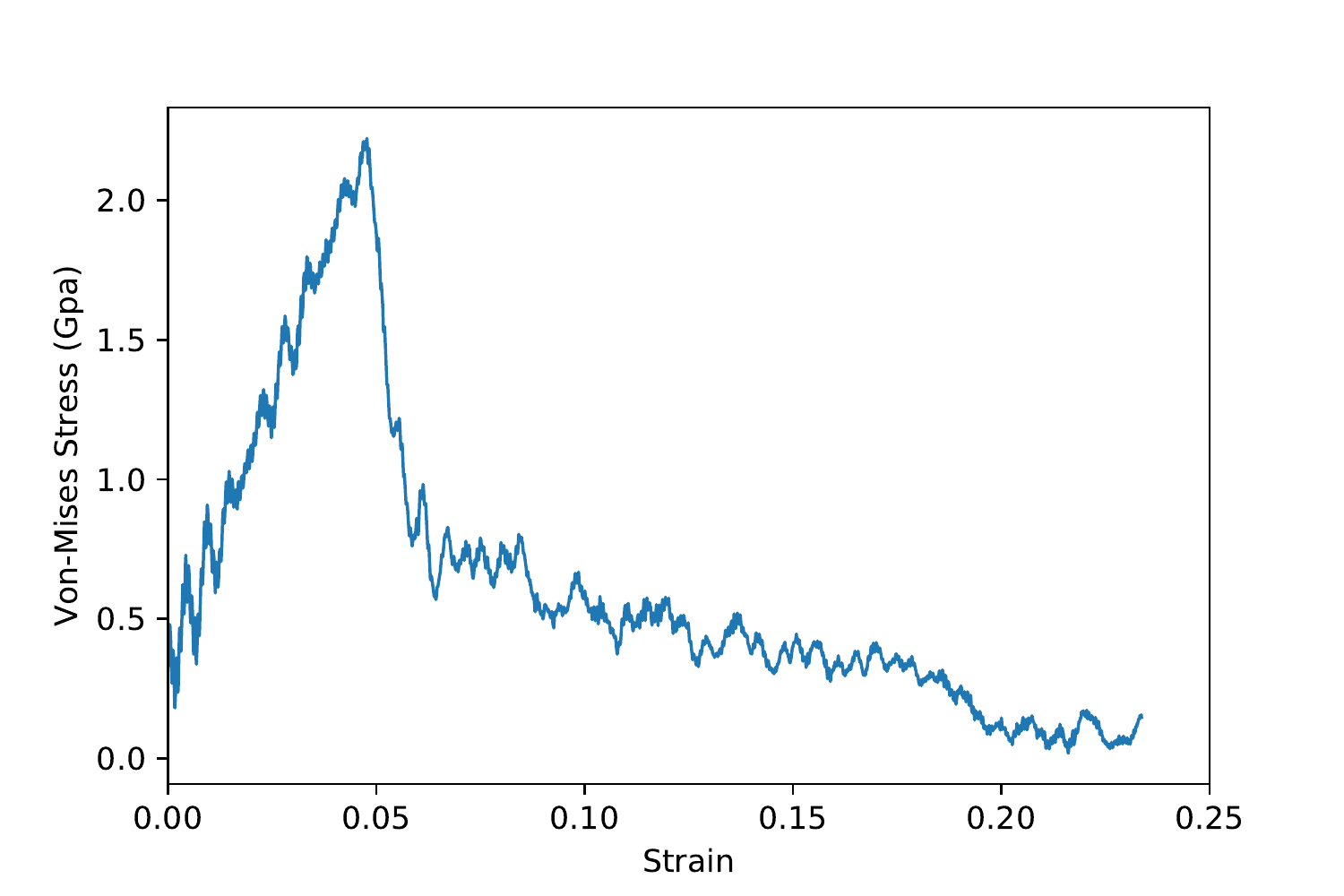}}
		\subfigure[with voids]{\includegraphics[width=2.3in]{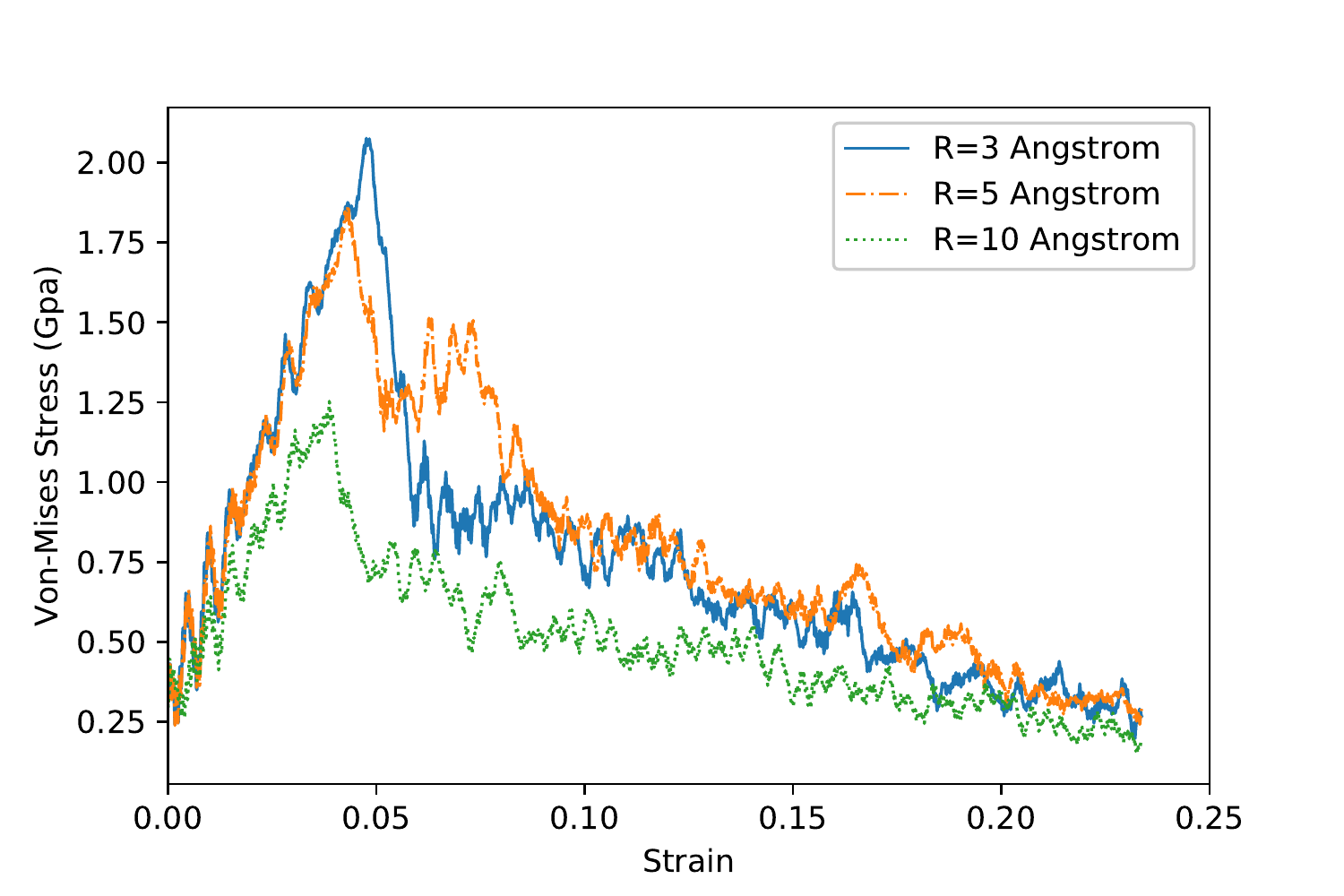}}\\
		
		\subfigure[with Cu inclusions]{\includegraphics[width=2.3in]{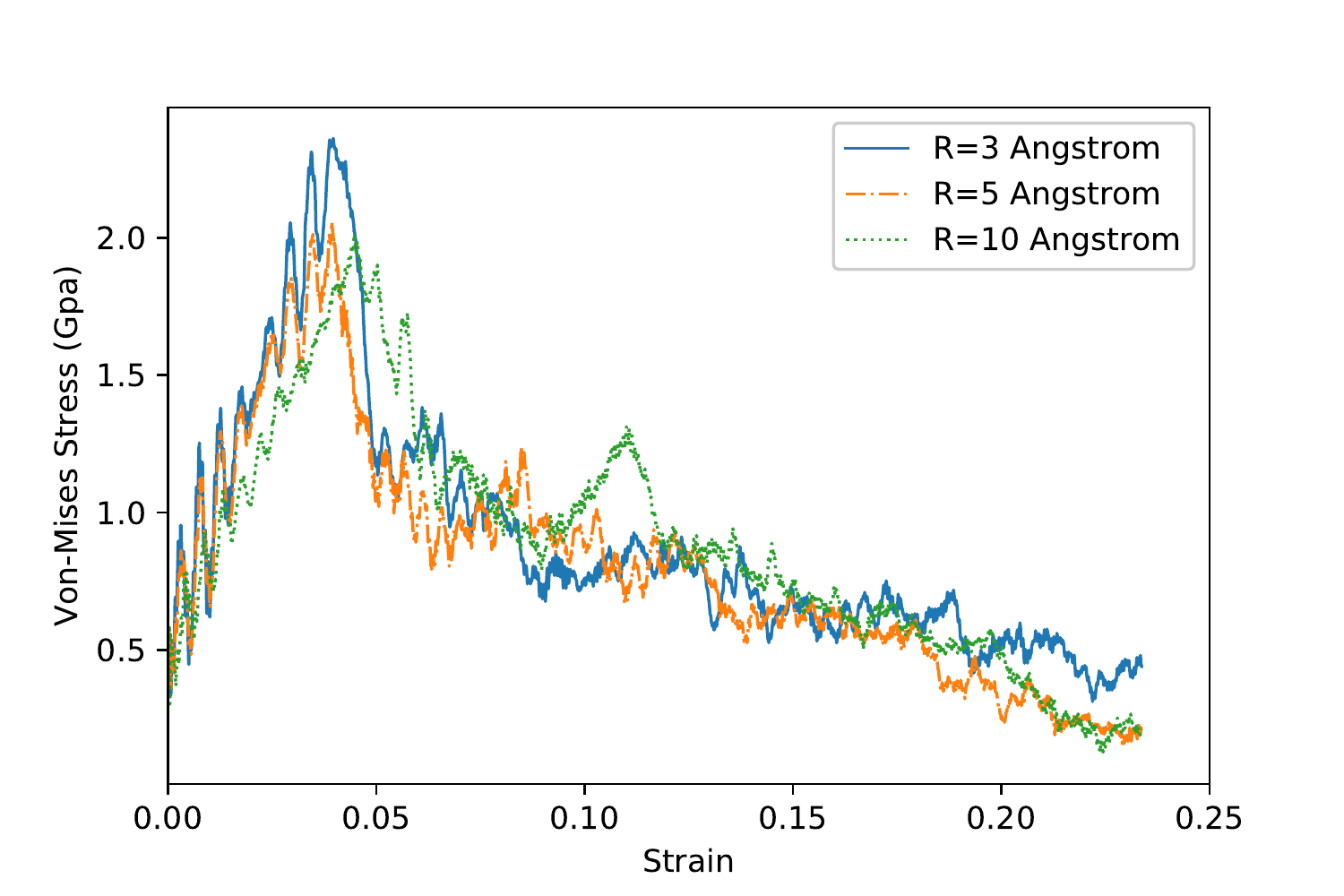}}
		\subfigure[with Al inclusions]{\includegraphics[width=2.3in]{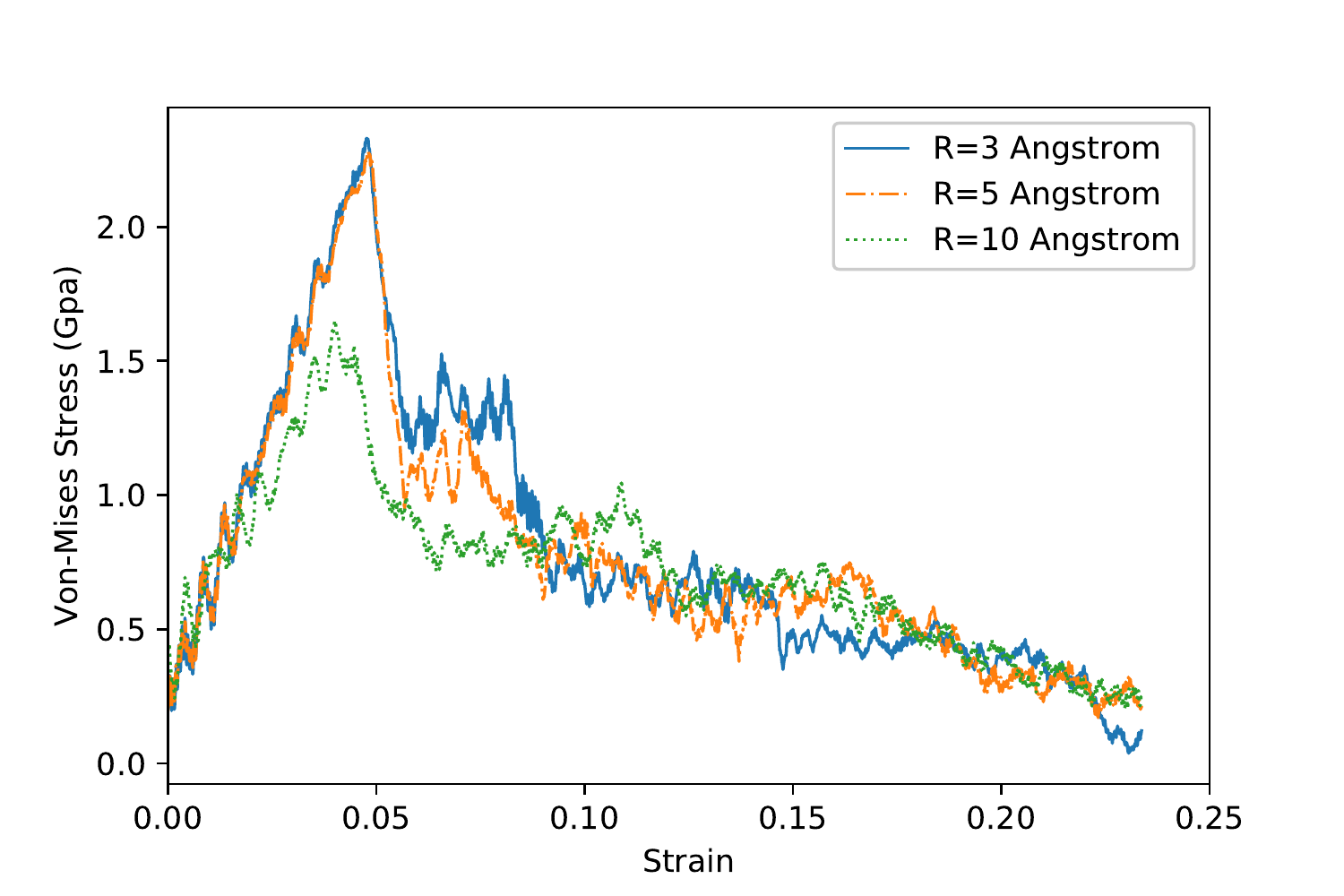}}
		\caption{Stress-strain curve for crack growing in a single crystal of nickel (a) without voids and inclusions, (b) with voids, (c) with Cu inclusions and (d) with Al inclusions.}
		\label{fig:strain_stress}
	\end{figure*}

	The above results showed the mechanical properties of crack propagation in a Ni crystal with voids and inclusions. In order to make comparisons between those results, the crack length was monitored during the crack propagation process. Figure \ref{fig:crack_l} shows the crack length as a function of the strain for crack propagation. It can be clearly seen that the crack starts growing at the strain of 5\%, then grows rapidly after the strain reaches 15\%  and finally the system fractures at the strain of 18\% as shown in Fig. \ref{fig:crack_l}a. Figure \ref{fig:crack_l}b shows that the crack grows rapidly after the strain reaches 16\% and the system fractures at the strain of 24\% for the model of a crystal with a void of $R_v=3$ \AA while the cracks just have slight increment even the strain reaches 24\% for the models with voids of $R_v=5$ \AA and $R_v=10$ \AA. Therefore, Fig. \ref{fig:crack_l}b indicates that the voids can absorb the strain energy, which will lead a better resistance to plastic deformation in crystals. Figure \ref{fig:crack_l}c and \ref{fig:crack_l}d also show that the inclusions can improve the resistance to plastic deformation since the systems fracture after the strain increases to 24\%. Figure \ref{fig:strain_stress} shows the stress-strain curves for crack propagation process. It can be clearly seen that the curves increase roughly linearly with the increasing strain before the stress reaches the first peak point which is about at the strain of 5 \%. Combined with Fig. \ref{fig:crack_l}, it can be determined that the cracks start growing after the stress reaches the first peak point. Therefore, the value of the stress at the first peak point can be treated as the critical stress for the process of crystal crack propagation. Moreover, Fig. \ref{fig:strain_stress} also shows that the voids decrease the critical stress of the mono-crystal while the inclusions have no significant effect on it.
	
	\section{conclusions}
	In this study, MD simulations were used to study the effects of voids and inclusions on the mechanical properties of crack propagation in a mono-crystal nickel. The distributions of crack tip stress, crack length and von Mises stress simulated by the models with different configurations of voids and inclusions were discussed. The main conclusions were summarized as follows:
	
	\begin{enumerate}[(1)]
		\item The path of crack propagation in a crystal can be changed by voids and inclusions.
		\item For the models of mono-crystals with voids, the results indicate that the voids can absorb the strain energy, which can lead a better resistance to plastic deformation in crystals, but the critical stress of the system should also be decreased by voids. 
		\item For the models of mono-crystals with inclusions, the results demonstrated that the inclusions could lead a better resistance to plastic deformation and different materials of inclusions have different performance. The results showed that both the crystals with inclusions of copper and aluminum required a higher strain to cause a fracture of the system.
	\end{enumerate}
	
	\section*{acknowledgements}
	This work was partially supported by Project of the Key Program of National Natural Science Foundation of China (Grant Number 11972155). This work is also partially supported by Hunan Provincial Innovation Foundation For Postgraduate (CX2018B203).

	\bibliographystyle{unsrt}
	\bibliography{manuscript}

\end{document}